\documentclass[mathematics,article,submit,pdftex,moreauthors]{Definitions/mdpi} 
\usepackage{natbib}
\usepackage{graphicx} 
\usepackage{caption}
\usepackage{subcaption} 
\usepackage{tabularx}
\usepackage{bm}
\usepackage{xcolor}
\usepackage{xspace}
\usepackage{multirow}
\usepackage{algorithm}
\usepackage{algorithmicx}
\usepackage{algpseudocode}

\firstpage{1} 
\pubvolume{1}
\issuenum{1}
\articlenumber{0}
\pubyear{2023}
\copyrightyear{2023}
\datereceived{ } 
\dateaccepted{ } 
\datepublished{ } 
\hreflink{https://doi.org/} 

 \Title{Studying Disease Reinfection Rates, Vaccine Efficacy and the Timing of Vaccine Rollout in the context of Infectious Diseases}

\TitleCitation{Studying Disease Reinfection Rates, Vaccine Efficacy and the Timing of Vaccine Rollout in the context of Infectious Diseases}

\Author{Elizabeth B. Amona$^{1}$\orcidA, Indranil Sahoo$^{1}$*\orcidB, Edward L. Boone$^{1}$ and Ryad Ghanam$^{2}$}

\AuthorCitation{Amona, E.; Sahoo, I.; Boone, E.; Ghanam, R.}

\address{%
$^{1}$ \quad Virginia Commonwealth University; amonaeb@vcu.edu; sahooi@vcu.edu; elboone@vcu.edu\\
$^{2}$ \quad Virginia Commonwealth University, Doha, Qatar; raghanam@vcu.edu}

\corres{Correspondence: sahooi@vcu.edu}

\abstract{
Qatar has undergone distinct waves of COVID-19 infections, compounded by the emergence of variants, posing additional complexities. This research uniquely delves into the varied efficacy of existing vaccines and the pivotal role of vaccination timing in the context of COVID-19. Departing from conventional modeling, we introduce two models that account for the impact of vaccines on infections, reinfections, and deaths. Recognizing the intricacy of these models, we use the Bayesian framework and specifically utilize the Metropolis-Hastings Sampler for estimation of model parameters. The study conducts scenario analyses on two models, quantifying the duration during which the healthcare system in Qatar could have potentially been overwhelmed by an influx of new COVID-19 cases surpassing the available hospital beds. Additionally, the research explores similarities in predictive probability distributions of cumulative infections, reinfections, and deaths, employing the Hellinger distance metric. Comparative analysis, employing the Bayes factor, underscores the plausibility of a model assuming a different susceptibility rate to reinfection, as opposed to assuming the same susceptibility rate for both infections and reinfections. Results highlight the adverse outcomes associated with delayed vaccination, emphasizing the efficacy of early vaccination in reducing infections, reinfections, and deaths. Our research advocates prioritizing early vaccination as a key strategy in effectively combating future pandemics. This study contributes vital insights for evidence-based public health interventions, providing clarity on vaccination strategies and reinforcing preparedness for challenges posed by infectious diseases.}


\begin{document}




\section{Introduction}\label{intro}
The relentless battle against the severe acute respiratory syndrome coronavirus 2 (SARS-CoV-2) continues to unveil new complexities in the dynamics of infectious diseases and the effectiveness of interventions. As the pandemic progresses, one aspect that has garnered significant attention is the phenomenon of reinfection — a topic both timely and imperative for further exploration. Reinfection with SARS-CoV-2 occurs when an individual contracts the virus, recovers, and subsequently becomes infected again. While most reinfections tend to be mild, severe illness can also occur \cite{willyard2023repeat, CDC}. People who are reinfected can also spread the virus to others, and staying up to date with vaccine doses and starting treatment within days after developing symptoms can decrease a person’s risk of experiencing severe illness from reinfection \cite{CDC}. 

The State of Qatar, like many countries, has experienced distinct waves of COVID-19 infections. From the initial surge of cases to the subsequent introduction of new variants, Qatar's experience serves as an insightful case study. In Qatar, an initial wave of infections occurred between March and June 2020, resulting in the development of detectable antibodies against SARS-CoV-2 in approximately $40\%$ of the population. Subsequently, the country experienced two consecutive waves of infections from January to May 2021, triggered by the emergence of the B.1.1.7 (alpha) and B.1.351 (beta) variants \citep{abu2021severity}. Other variants such as Omicron and its subvariants have also emerged \citep{chemaitelly2021waning}. Understanding the patterns of infections, reinfections, and associated mortality in the presence of effective vaccination is crucial for guiding effective public health strategies. To this end, this research presents a comprehensive methodology to examine the impact of variations in vaccine efficacy and the timing of vaccine administration in the context of infectious diseases. The proposed methodology has been implemented to analyze the dynamics of COVID-19 in terms of infections, reinfections, and deaths within Qatar.

Various studies have investigated the impact of vaccination effectiveness on infections and deaths globally, employing mathematical frameworks or case-control designs. \cite{amona2023incorporating} developed a mathematical (SEIRDV) model to study the effectiveness of vaccination in the reduction of secondary cases and mortality rate in Qatar. The authors concluded that vaccines drastically reduced the basic reproduction number, $R_0$ and saved lives. \cite{ghosh2023mathematical} employed an SEIR model to determine the required vaccine efficiency to diminish infection and mortality peaks in Italy, India, and Australia. Results indicated substantial reductions with specific vaccine efficiency and coverage combinations; for example, a 75\% efficient vaccine administered to 50\% of the population can reduce the peak number of infected individuals by nearly 50\% in Italy. \citep{andrews2022covid} conducted a test-negative case–control study in England to estimate vaccine effectiveness against omicron and delta variants, revealing effectiveness against symptomatic disease, but highlighting waning protection over time. Some studies have further delved into the possibility of reinfection after the waning of the vaccines. \cite{hammerman2022effectiveness} performed a retrospective cohort study in Israel, revealing a sixfold higher reinfection rate among the unvaccinated compared to the vaccinated, highlighting vaccine effectiveness and waning protection. Additional cohort studies on vaccine effectiveness and reinfection risks post-waning are documented in \cite{hiam2022waning, lewis2022effectiveness, hall2022protection, sheehan2021reinfection, rahman2022covid}. Notably, these studies exclusively employed cohort designs. \cite{mukandavire2020quantifying} explored scenario analysis early in the pandemic using an SEIR model in South Africa in a different context. 

Thus, after an extensive literature review, several research gaps in the context of studying the dynamics of infectious diseases have emerged. Firstly, timely vaccine availability is identified as a significant concern that requires further investigation. Secondly, there is a notable absence of mathematical models that explicitly explain the degree of susceptibility after infection resulting from the waning of vaccination. The lack of such models leave a critical gap in understanding the dynamics of post-vaccination susceptibility. Thirdly, a gap also exists in the realm of scenario analysis on vaccine efficacy. No study has comprehensively explored how variations in vaccine efficacy might impact infection, reinfection, and mortality rates. This gap underscores the need for research that delves into the broader implications of vaccine efficacy on different aspects of disease dynamics.

The uniqueness of our study lies in its emphasis on scenario analysis of vaccine efficacy, closely reflecting the real-world dynamics of vaccination efforts. In this context, we explore the impact of these variations on the spread of the virus, as well as the occurrence of reinfections and mortality. In conjunction with vaccine efficacy, the timing (early or late) of vaccine roll out is identified as an important factor. The choice between early and late vaccine administration holds implications for achieving population-level immunity and attenuating the impact of the virus. To better reflect real-world scenarios, we propose two mathematical models with different dynamics and under different assumptions (see Section \ref{models}). This work deviates from traditional mathematical modeling by not only estimating parameters and studying system dynamics based on real data, but also projecting various scenarios relevant to infectious diseases, specifically COVID-19. Furthermore, while conventional models change parameter values to study the efficacy and impact of vaccination, we are changing when the interventions and/or vaccinations were implemented to explore the dynamics, which makes our study novel. Given our proposed models and the estimated parameter values, we seek to address the following research questions: What would have happened if vaccines were administered at different time points, either earlier or later? In scenarios with limited government interventions, what potential consequences might arise? Is there a possibility that delaying vaccine administration could lead to a scenario where the number of infections exceeds the capacity of the country's healthcare system, specifically in terms of available hospital beds?? We aim to calculate the number of days on which such system breakdowns might occur, should vaccines be administered late. Thus, our objective is to utilize the proposed mathematical models as a tool to answer the above research questions and understand and interpret the consequences of these different strategies, rather than just parameter estimation or data fitting. 

\subsection{Motivation for model comparison}

In order to explain the real-world dynamics of infectious diseases, we introduce two mathematical models, each offering a different viewpoint on the complex interaction among diseases, immunity, and vaccination. The first model assumes that individuals who recover from their initial infection might exhibit different susceptibility rates upon re-exposure. This assumption reflects the idea that immunity wanes over time and that those who have recovered from the disease may still be at risk, although at a different rate before reinfection. This model thus considers different susceptibility rates among different segments of the population, and their impact on the overall course of the disease.

In contrast, the second model does not account for the varying susceptibility of those who have recovered, thereby taking a more straightforward approach and assuming a uniform susceptibility rate among all individuals. We describe the two models in Section \ref{models}, thereby exploring the question: Does the first model, with its complex assumptions about variable susceptibility, represent real-world dynamics better? Or does the simplicity of the second model give a more accurate picture? Thus, our approach gives data-driven insights into the real-world implications of varying susceptibility rates, immunity waning, and vaccination efficacy. As such, this study provides a clearer understanding of disease dynamics in the presence of evolving immunity, and guidance for public health interventions and decision-making in a complex, ever-changing world.

\section{The Compartmental Models} \label{models}

The models presented in this paper are an extension of the SEIRDV model as introduced by \cite{amona2023incorporating}. Hence, all assumptions inherent to the SEIRDV model remain applicable to our proposed models. In these models, $t$ serves as a time index, denoting the number of days elapsed since the first recorded case of COVID-19 within the population of interest. At any given time $t$, $S(t)$ denotes the total number of susceptible individuals, $S_1(t)$ represents the number of individuals who are susceptible and have not contracted the disease, $E(t)$ denotes the number of individuals who are in the exposed state, $I(t)$ denotes the total number of individuals who are infected (displaying symptoms), $R_E (t)$ denotes the cumulative number of individuals who have recovered following exposure, $R_I(t)$ denotes the cumulative number of individuals who have recovered following infection, $D(t)$ denotes the cumulative number of deaths, $S_2 (t)$ denotes the cumulative number of individuals who have regained susceptibility after their initial infection, $I_I (t)$ denotes the cumulative number of individuals who have experienced reinfection, $R_R (t)$ denotes the cumulative number of individuals who have recovered following reinfection, and $V(t)$ denotes the cumulative number of vaccinated individuals.

\subsection{The $S_1$EIRDV$S_2 I_I R_R$ Model }\label{model1}
\begin{figure}
\begin{center}
\includegraphics[width = 0.8\textwidth]{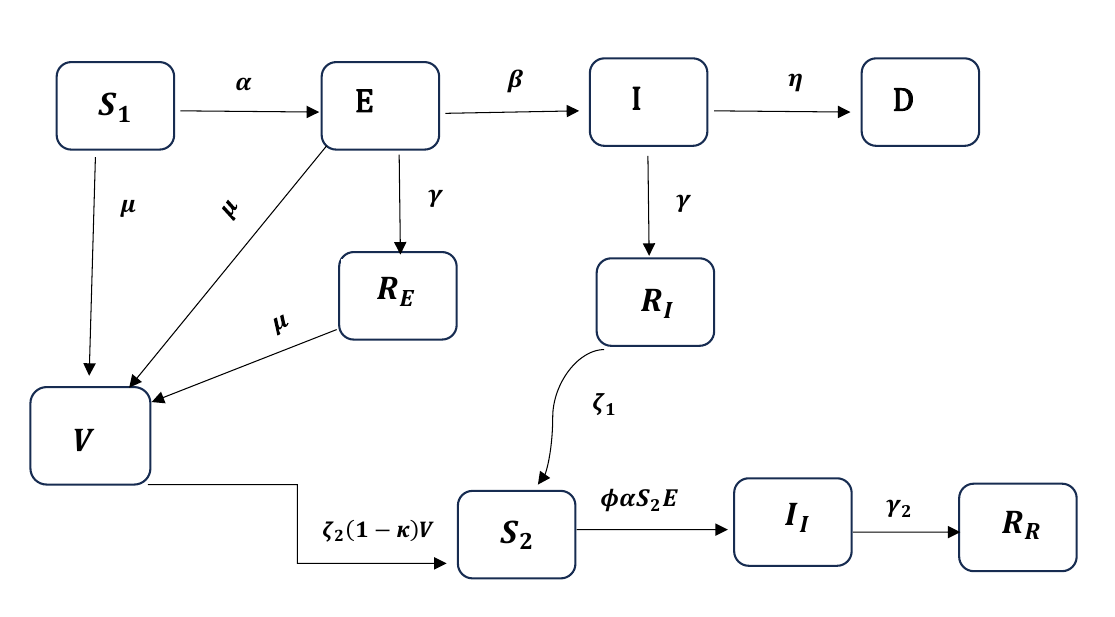} 
\caption{Schematic diagram of $S_1$EIRDV$S_2 I_I R_R$ model for Covid-19\label{fig:Schematic2}}
\end{center}
\end{figure}

In addition to the inherent model assumptions, we assume that individuals who have previously been infected would develop antibodies for a certain period before these antibodies naturally wane. Subsequently, these individuals transition to a different Susceptible compartment ($S_2$) at a different rate compared to the initial Susceptible compartment ($S_1$). Additionally, we assume that the waning of natural immunity and immunity resulting from vaccination occur at different rates, denoted as $\zeta_1$ (representing natural immunity waning) and $\zeta_2 (1-\kappa)$ (representing vaccination-induced waning). Moreover, we assume that all individuals who experience reinfection would recover and transition to a secondary recovery compartment ($R_R$). We do not consider any deaths following reinfection. This assumption is supported by research indicating that the risk of severe reinfection in the State of Qatar was only approximately 1\% of the risk faced by individuals who had not been previously infected \cite{abu2021severity}. The compartmental model $S_1$EIRDV$S_2 I_I R_R$ thus comprises of ten distinct compartments, and a visual representation is presented in Figure \ref{fig:Schematic2}. 

Following the flow diagram given in Figure \ref{fig:Schematic2} and the modeling assumptions described above, Model 1 can be formally represented by the following system of equations:
 
 \begin{eqnarray}
\label{eq:Sys1}
\frac{dS_1}{dt} &=& -\alpha S_1(t) E(t) - \mu S_1(t), \nonumber\\ 
\frac{dE}{dt} &=& \alpha S_1(t) E(t) - (\beta + \gamma_1 + \mu) E(t), \nonumber\\
\frac{dI}{dt}  &=&   \beta E(t)  - (\gamma_1 + \eta)I(t) , \nonumber\\
\frac{dR_E}{dt}  &=&  \gamma_1 E(t) - \mu R_E(t), \nonumber\\
\frac{dR_I}{dt}  &=&  \gamma_1 I(t)  - \zeta_1 R_I(t), \\
\frac{dD}{dt}  &=&  \eta I(t) , \nonumber\\
\frac{dS_2}{dt}  &=&  \zeta_1 R_I(t)  - \phi \alpha S_2(t) E(t) + \zeta_2 (1-\kappa) V(t), \nonumber\\
\frac{dI_I}{dt}  &=&  \phi \alpha S_2(t) E(t) - \gamma_2 I_I(t), \nonumber\\
\frac{dR_R}{dt}  &=&   \gamma_2  I_I(t), \nonumber\\
\frac{dV}{dt}  &=&  \mu (S_1(t) +  E(t) +  R_E(t)) - \zeta_2 (1-\kappa) V(t), \nonumber
\end{eqnarray}
with the following constraints $~S_1(t) \geq 0, ~S_2(t) \geq 0, ~ E(t) \geq 0, ~I(t) \geq 0, ~ I_I(t) \geq 0, ~  R_E(t) \geq 0, ~ R_I(t) \geq 0, ~ R_R(t) \geq 0, ~D(t) \geq 0, ~ \text{and} ~V (t) \geq 0$. All parameters in system (\ref{eq:Sys1}) are positive and are explained as follows: $\alpha$ is the transmission rate (per day $\times$ individual$^2$) from Susceptible to Exposed, $\beta$ denotes the rate (per day) at which Exposed become Infected (symptomatic), $\gamma_1$ is the rate (per day) at which Infected become Recovered, $\phi$ is the rate (per day $\times$ individual$^2$) at which those who recovered from the first infections become reinfected, $\gamma_2$ is the rate (per day) at which the Reinfected become Recovered, the vaccination rate (per day) is denoted by $\mu$, the efficacy of the vaccine is denoted by $\kappa$, where $0 < \kappa \leq 1$, the waning of the natural immunity and vaccine immunity are denoted by $\zeta_1$ and $\zeta_2$ respectively, and the mortality rate (per day) for those Infected is denoted by $\eta$. Note that $\gamma_2 = 1$, which means that everyone who got reinfected recovers after a while, as a consequence of one of our assumptions.

\subsection{The SEIRDV$I_I R_R$ Model}\label{model2}
\begin{figure}
\begin{center}
\includegraphics[width = 0.7\textwidth]{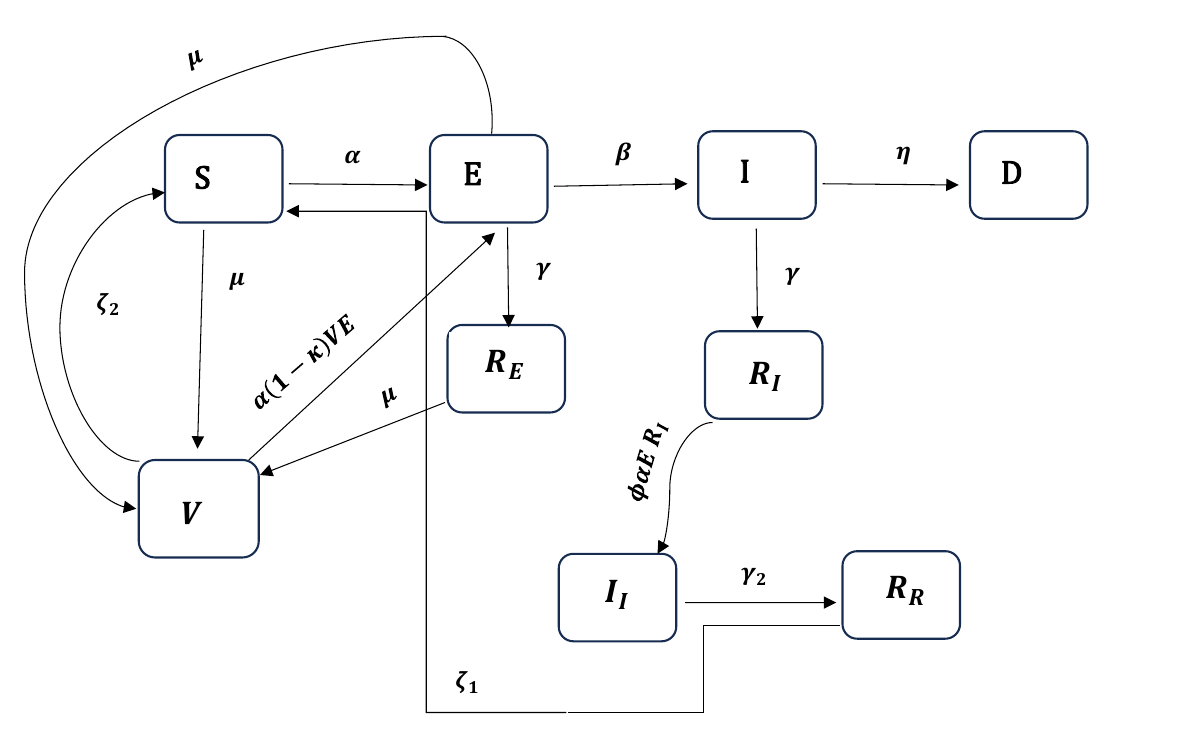} 
\caption{Schematic diagram of SEIRDV$I_I R_R$ model for Covid-19\label{fig:Schematic}}
\end{center}
\end{figure}

In this model, we assume that individuals who recover after the first infection could become reinfected within a short period of time if they come in contact with exposed individuals. In Model \ref{model2}, we do not have the second susceptible compartment ($S_2$) as we had in Model \ref{model1}. This means that when an individual recovers, they would automatically move into the reinfected compartment after the natural immunity wanes, as opposed to Model \ref{model1}, which assumes that there is a delay between recovery and reinfections through the second susceptible compartment ($S_2$) and that the susceptibility rate differs. The remaining assumptions of this model are similar to the previous model. A visual representation is presented in Figure \ref{fig:Schematic}.

Following the flow diagram given in Figure \ref{fig:Schematic} and the modeling assumptions described above, we represent Model 2 using the following system of equations:

\begin{eqnarray}
\label{eq:Sys2}
\frac{dS}{dt} &=& -\alpha S(t) E(t) - \mu S(t) + \zeta_1 R_R(t)  + \zeta_2 V(t), \nonumber\\
\frac{dE}{dt} &=& \alpha S(t) E(t) - (\beta + \gamma_1 + \mu) E(t) + \alpha (1-\kappa)V(t)E(t), \nonumber\\
\frac{dI}{dt} &=&  \beta E(t)  - (\gamma_1 + \eta)I(t) , \nonumber\\
\frac{dR_E}{dt} &=& \gamma_1 E(t) - \mu R_E(t), \nonumber \\
\frac{dR_I}{dt} &=& \gamma_1 I(t)  - \alpha \phi E(t) R_I(t),\\
\frac{dD}{dt} &=& \eta I(t) , \nonumber\\
\frac{dI_I}{dt} &=& \alpha \phi E(t) R_I(t)  - \gamma_2 I_I(t), \nonumber\\
\frac{dR_R}{dt} &=&  \gamma_2 I_I(t) - \zeta_1 R_R (t), \nonumber\\
\frac{dV}{dt} &=& \mu (S(t) +  E(t) +  R_E(t)) - \zeta_2 V (t) - \alpha (1-\kappa)V(t)E(t) , \nonumber
\end{eqnarray}
with the following constraints $~S(t) \geq 0, ~ E(t) \geq 0, ~I(t) \geq 0, ~ I_I(t) \geq 0, ~  R_E(t) \geq 0, ~ R_I(t) \geq 0, ~ R_R(t) \geq 0, ~D(t) \geq 0, ~ \text{and} ~V (t) \geq 0$. All model parameters can be interpreted similarly as those discussed in subsection \ref{model1}.

\section{Statistical Methodology for Model Inference and Model Comparison}\label{method}
\subsection{The Bayesian Analysis Framework}

Given the evolving nature of the pandemic and the interventions by governments to influence various parameters, assuming a ``steady state'' for the dynamic system is inappropriate. Hence, we allow the transmission, recovery and reinfection rates in our models to vary over time. We define the transmission rate vector as $\bm{\alpha} = \left( \alpha_0, \alpha_1,...\alpha_m \right)^{T}$, where $\alpha(t) = \alpha_{m - 1}  \text{ if } t_{m-1} \leq t < t_{m}$. Since $\alpha(t) > 0$ for all $t$, we have $\alpha_i > 0, i = 0, 1, \ldots, m$. Similarly, we define the recovery rate vector as $\bm{\gamma}_1 = (\gamma_{10}, \gamma_{11}, \cdots, \gamma_{1l})^T, l \leq m$, where each of the $\gamma$'s are independent and denotes the changed recovery rate once an intervention has been administered. We write $\gamma_1 (t) = \gamma_{1 \hspace{0.1cm} l - 1}, \text{ if } t_{l-1}' \leq t < t_{l}'$, where $t_1', t_2', \ldots, t_l' \in \lbrace t_1, t_1 + 1, t_1 +2, \ldots, t_2, t_2 + 1, \ldots, t_m \rbrace$. Since $\gamma_1(t) > 0$ for all $t$, we have $\gamma_{1j} > 0, j = 0, 1, \ldots, l$. Similarly, the reinfection rate vector is written as 
$\bm{\phi} = \left( \phi_0, \phi_1,...\phi_p \right)^{T}$, where $\phi(t) = \phi_{p - 1}  \text{ if } t_{p - 1}'' \leq t < t_{p}''$ with $\phi_k > 0, k = 0, 1, \ldots, p$ and $t_1'', t_2'', \ldots, t_p'' \in \lbrace t_1, t_1 + 1, t_1 +2, \ldots, t_2, t_2 + 1, \ldots, t_m \rbrace$.

Thus, for $i = 0, 1, \ldots, m, j = 0, 1, \ldots, l$ and $k = 0, 1, \ldots, p$, the system of equation in \ref{eq:Sys1} becomes

 \begin{eqnarray}
\label{eq:Sys3}
\frac{d\lambda_{S_1}}{dt} &=& - \alpha_i \lambda_{S_1}(t) \lambda_{E}(t) - \mu \lambda_{S_1}(t), \nonumber\\ 
\frac{d \lambda_E}{dt} &=& \alpha_i \lambda_{S_1}(t) \lambda_E(t) - (\beta + \gamma_{1j} + \mu )\lambda_E(t), \nonumber\\
\frac{d\lambda_I}{dt}  &=&   \beta \lambda_E(t)  - (\gamma_{1j} + \eta )\lambda_I(t)  , \nonumber\\
\frac{d\lambda{R_E}}{dt}  &=&  \gamma_{1j} \lambda_E(t) - \mu \lambda_{R_E}(t), \nonumber\\
\frac{d\lambda_{R_I}}{dt}  &=&  \gamma_{1j} \lambda_I(t)  - \zeta_1 \lambda_{R_I}(t), \\
\frac{d\lambda_D}{dt}  &=&  \eta \lambda_I(t) , \nonumber\\
\frac{d\lambda_{S_2}}{dt}  &=&  \zeta_1 \lambda_{R_I}(t)  - \phi_k \alpha_i \lambda_{S_2}(t) \lambda_E(t) + \zeta_2 (1-\kappa) \lambda_V(t), \nonumber\\
\frac{d\lambda_{I_I}}{dt}  &=&  \phi_k \alpha_i \lambda_{S_2}(t) \lambda_E(t) - \gamma_2 \lambda_{I_I}(t), \nonumber\\
\frac{d\lambda_{R_R}}{dt}  &=&   \gamma_2  \lambda_{I_I}(t), \nonumber\\
\frac{d\lambda_V}{dt}  &=&  \mu (\lambda_{S_1}(t) +  \lambda_E(t) +  \lambda_{R_E}(t)) - \zeta_2 (1-\kappa) \lambda_V(t), \nonumber
\end{eqnarray}

Similarly, the system of equation in \ref{eq:Sys2} becomes

 \begin{eqnarray}
\label{eq:Sys4}
\frac{d\lambda_{S_1}}{dt} &=& - \alpha_i \lambda_{S_1}(t) \lambda_{E}(t) - \mu \lambda_{S_1}(t) + \zeta_1 \lambda_{R_R}(t) + \zeta_2 \lambda_V(t), \nonumber\\ 
\frac{d \lambda_E}{dt} &=& \alpha_i \lambda_{S_1}(t) \lambda_E(t) - (\beta + \gamma_{1j} + \mu ) \lambda_E(t) + \alpha_i (1 - \kappa) \lambda_V(t) \lambda_E(t), \nonumber\\
\frac{d\lambda_I}{dt}  &=&   \beta \lambda_E(t)  - ( \gamma_{1j} + \eta ) \lambda_I(t)  , \nonumber\\
\frac{d\lambda{R_E}}{dt}  &=&  \gamma_{1j} \lambda_E(t) - \mu \lambda_{R_E}(t), \nonumber\\
\frac{d\lambda_{R_I}}{dt}  &=&  \gamma_{1j} \lambda_I(t)  - \alpha_i \phi_k \lambda_E(t) \lambda_{R_I}(t), \\
\frac{d\lambda_D}{dt}  &=&  \eta \lambda_I(t) , \nonumber\\
\frac{d\lambda_{I_I}}{dt}  &=&   \alpha_i \phi_k \lambda_E(t) \lambda_{R_I}(t) - \gamma_2 \lambda_{I_I}(t), \nonumber\\
\frac{d\lambda_{R_R}}{dt}  &=&   \gamma_2  \lambda_{I_I}(t) - \zeta_1 \lambda_{R_R}(t), \nonumber\\
\frac{d\lambda_V}{dt}  &=&  \mu (\lambda_{S}(t) +  \lambda_E(t) +  \lambda_{R_E}(t) ) - \zeta_2 \lambda_V(t) - \alpha_i (1 - \kappa) \lambda_V(t) \lambda_E(t), \nonumber
\end{eqnarray}
where $\lambda_{S_1}(t), \lambda_E(t), \lambda_I(t), \lambda_{R_E}(t), \lambda_{R_I}, \lambda_{S_2}(t), \lambda_V(t), \lambda_{I_I}(t), \lambda_{R_R}(t)$ and $\lambda_D(t)$ denote the mean parameters, respectively. The prior distributions of parameters in models \ref{eq:Sys3} and \ref{eq:Sys4} are chosen as
\begin{equation}\label{eq:prior}
\begin{aligned}
\alpha_i &\sim Exp(1), \hspace{0.25cm} i = 0, 1, \ldots, m = 10 ,\cr
\beta &\sim Exp(1) ,\cr
\gamma_{1j} &\sim Exp(1), \hspace{0.25cm} j = 0, 1,  \ldots, l = 3 ,\cr
\zeta_1 &\sim Exp(1) ,\cr
\zeta_2 &\sim Exp(1) ,\cr
\mu &\sim Exp(1) ,\cr
\phi_k &\sim Exp(1) ,\hspace{0.25cm} k = 0, 1,  \ldots, p = 3,\cr
\gamma_2 &\sim Exp(1),\cr
\eta &\sim Exp(1) ,\cr
\kappa &\sim Be(1,1) ,\cr
\end{aligned}                            
\end{equation}
The likelihoods for $I(t)$, $R_I(t)$, $I_I(t)$, $R_R(t)$, $D(t)$ and $V(t)$ in models \ref{eq:Sys3} and \ref{eq:Sys4} are given by
\begin{equation}\label{eq:Like1}
\begin{aligned}
I(t) &\sim Poisson \left( \lambda_I(t) \right) ,\cr
R_I(t) &\sim Poisson \left( \lambda_{R_I}(t) \right) ,\cr
D(t) &\sim Poisson \left( \lambda_D(t) \right) ,\cr
I_I(t) &\sim Poisson \left( \lambda_{I_I}(t) \right) ,\cr
R_R(t) &\sim Poisson \left( \lambda_{R_R}(t) \right) ,\cr
V(t) &\sim Poisson \left(\lambda_V(t) \right) .
\end{aligned}
\end{equation}

Note that  $S_1(t)$, $S_2(t)$, and $E(t)$ in model \ref{eq:Sys3} and $S(t)$, $E(t)$ in model \ref{eq:Sys4} are not in the likelihood as they are latent states in that they are not directly observed. While the true likelihood for $\{S_1(t), S_2 (t), E(t), I(t), R_E(t), R_I(t), I_I(t), R_R(t), V(t), D(t)\}$ should be multinomial, with three latent states in \ref{eq:Sys3} and two latent states in \ref{eq:Sys4}, one of which is the largest state, the multinomial approach is challenging to apply. Hence, this work adopts Poisson likelihood as an approximation. The posterior distribution for both models can be calculated based on the prior and likelihood specifications mentioned above.  Since it is difficult to solve the posterior distribution analytically, we employ the  Markov chain Monte Carlo (MCMC) technique to sample from the posterior distribution \citep{gelman1995bayesian}. Specifically, we utilized the Metropolis-Hastings sampler \citep{gilks1995markov, albert2009introduction, amona2023incorporating}.  

To tune the sampler, a series of short chains were generated and analyzed for convergence and adequate acceptance rates.  These initial short chains were discarded as ``burn-in'' samples. The tuned sampler was used to generate 50,000 samples from the posterior distribution and trace plots were visually examined to ensure convergence (not included).  All inferences, including parameter estimation and uncertainty quantification, were made from these 50,000 posterior draws. The model and sampling algorithm were custom programmed in the R statistical programming language version 3.6.3. The computations take approximately 10,800 seconds using an AMD A10-9700 3.50GHz processor with 16GB of RAM to obtain 50,000 draws from the posterior distribution. For more details on statistical inference see \cite{wackerly2014mathematical, casella2021statistical, berger1985prior}.

\subsection{Bayesian Model Comparison Using Bayes Factor}\label{bayes}

 In Bayesian statistics, model comparison is a fundamental task that allows researchers to evaluate the relative support for different hypotheses or models using available data. Particularly, in Bayesian model comparison, the Bayes Factor (BF) is a robust and versatile tool that helps researchers evaluate the relative probability of different models. By considering both parameter uncertainty and prior information, it offers a comprehensive and principled approach to model selection. When working with posterior samples obtained through methods like the Metropolis-Hastings sampler, Bayes Factor can be especially valuable for assessing the probability of competing models and informing decision-making in a wide range of scientific disciplines.

Furthermore, the Bayes Factor quantifies the support for one statistical model (Model 1) over another (Model 2) given the observed data. That is, the Bayes Factor offers a formal framework to test the following hypothesis based on the principles of Bayesian probability theory \cite{jeffreys1998theory}. 
\begin{align*}
\text{Null Hypothesis} (H_0) &: \text{Model 1 is more probable than Model 2. }\\
\text{Alternative Hypothesis} (H_A)&: \text{Model 2 is more probable than Model 1.}
\end{align*}
The Bayes Factor is computed as the ratio of the marginal likelihoods of the two models:
\[BF_{1 2} = \frac{P(Data | Model\, 1)}{P(Data | Model\, 2)} ,\]
\noindent where, \(P(Data | Model\, 1)\) is the marginal likelihood of Model $1$ and \(P(Data | Model\, 2)\) is the marginal likelihood of Model $2$. If the Bayes Factor is greater than 1, it indicates that Model $1$ is favored over Model $2$. The magnitude of $BF_{1 2}$ provides a measure of the strength of this preference. Larger $BF_{1 2}$ values indicate stronger evidence in favor of Model $1$. Conversely, Bayes Factor less than 1 suggests that Model $2$ is favored over Model $1$. A Bayes Factor close to 1 suggests that both models are equally plausible based on the data, thereby providing no strong evidence in favor of one model over the other \cite{lesaffre2012bayesian, kass1995bayes}. In our work, we compare the two models using Bayes Factor to provide a data-driven assessment of which model better represents the dynamics of the pandemic in the State of Qatar over our study period of February 28, 2020 - August 29, 2021. 

\subsection{Assessing `Closeness' of Density Plots using Hellinger Distance}

Density plots are powerful tools for visualizing and comparing probability distributions, particularly in the context of Bayesian analysis. While visual inspection of density plots can offer initial insights, quantifying the degree of `closeness' between two density plots is often necessary for rigorous statistical analysis. Among several metrics such as Kullback-Leibler Divergence, Jensen-Shannon Divergence and many more, the Hellinger distance is a valuable tool for assessing the similarity or dissimilarity of density plots and enhancing the interpretability of Bayesian posterior distributions. This metric is useful for quantifying the separation or overlap between probability density functions, among other characteristics, and it offers a robust and objective approach to comparing and evaluating the results of Bayesian analyses. While Kullback-Leibler divergence is usually a more popular choice, it is not symmetric \citep{goutis1998model} and hence is not suitable for our purposes. The Hellinger distance is defined as 
\[
H{^2}(f, g) = \frac{1}{2} \int{ \left(\sqrt{f(x)} - \sqrt{g(x)}\right)^2 dx}  =  1 - \int{\sqrt{f(x)g(x)} dx,} ~~~~~~~~ 0 \leq H^{2}(f,g) \leq 1.\]
\noindent where,  \(f(x)\) and \(g(x)\) are the two probability density functions being compared. The integral is computed over the entire support of the distributions \cite{goldenberg2019survey, birge1986estimating}. The Hellinger distance produces a value between 0 and 1, where \(H^2 = 0\) indicates that the two density plots are identical,  \(0 < H^2 < 1\) implies that the density plots share some degree of similarity, and \(H^2 = 1\) suggests that the density plots are completely dissimilar with no overlap \cite{boone2014hellinger}. 


In our case, \(f(x)\) and \(g(x)\) are the cumulative posterior predictive distributions, estimated using kernel density smoothing (KDE) with a Gaussian kernel and default bandwidth as suggested by \cite{scott2015multivariate}. The smoothing is implemented using the R function \texttt{density}() from the \texttt{stats} package.  Since KDEs can extend beyond the original data range, while comparing two KDEs, we ensure they (i) cover a common range of values and (ii) discretize that range identically, using an equally-spaced set of points in the analysis. The implementation details can be found in the 'ScenarioHellingerDistAnalysis.R' script at \url{https://github.com/elizabethamona/VaccinationTiming}. Since we focuse only on infected, reinfected, and deaths in our scenario analysis (see \ref{scen12}), we smooth out the cumulative posterior predictive densities for these three states for all six scenarios using the Gaussian kernel density.


\section{Results}\label{res}

In this section, we determine which of the two mathematical models introduced earlier more accurately reflects real-world dynamics of the COVID-19 pandemic. Specifically, we investigate the validity of the assumption that individuals recovering from initial infections transition to a second susceptible compartment at a different rate than those in the initial susceptible compartment. This assumption is explored to shed light on the complexities of understanding the impact of vaccination in the context of partial interventions. Notably, the second model omits consideration of the possibility that recovered individuals might be susceptible to reinfection at a different rate. 

\begin{figure}[h]
 \begin{subfigure}{0.49\textwidth}
     \includegraphics[width=\textwidth]{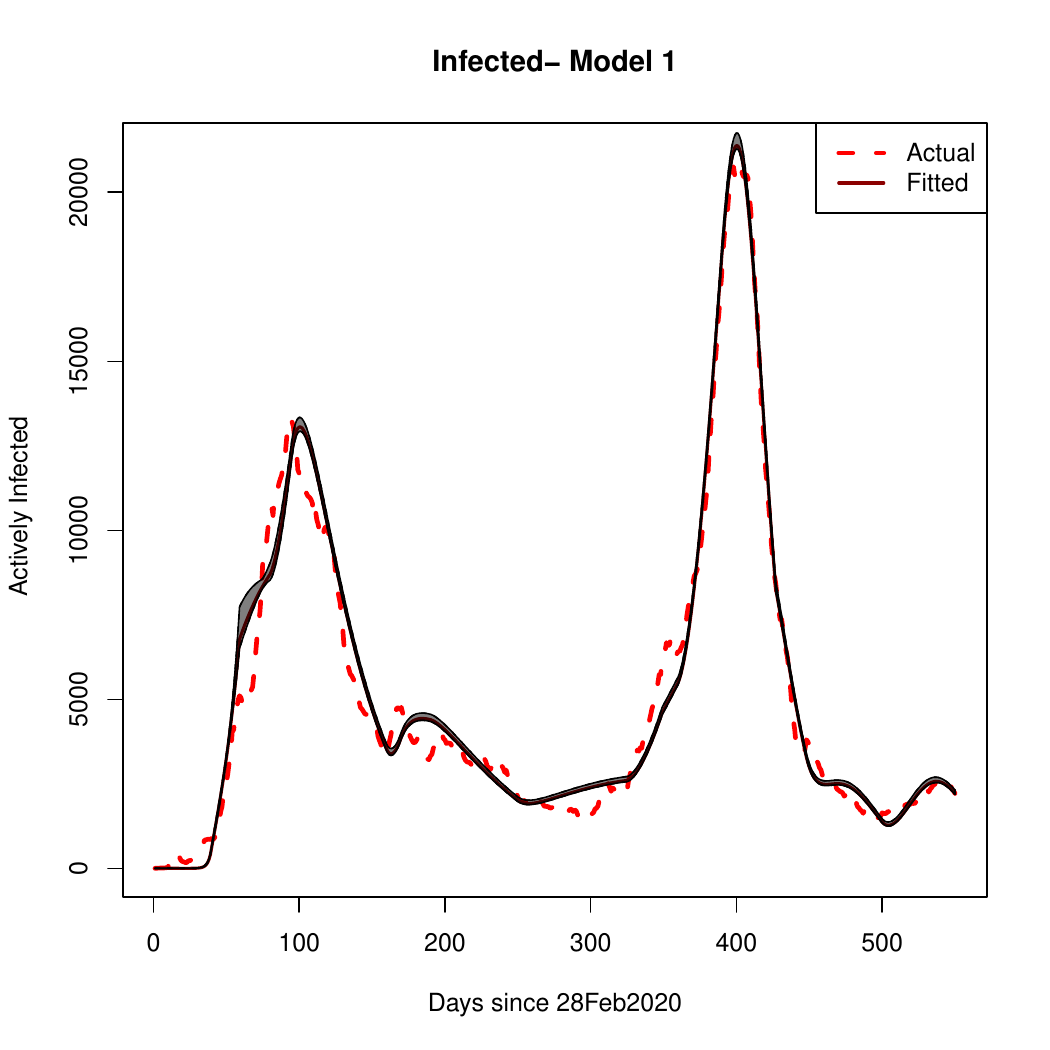}
     \caption{}
     \label{fig:a1}
 \end{subfigure}
 \hfill
 \begin{subfigure}{0.49\textwidth}
     \includegraphics[width=\textwidth]{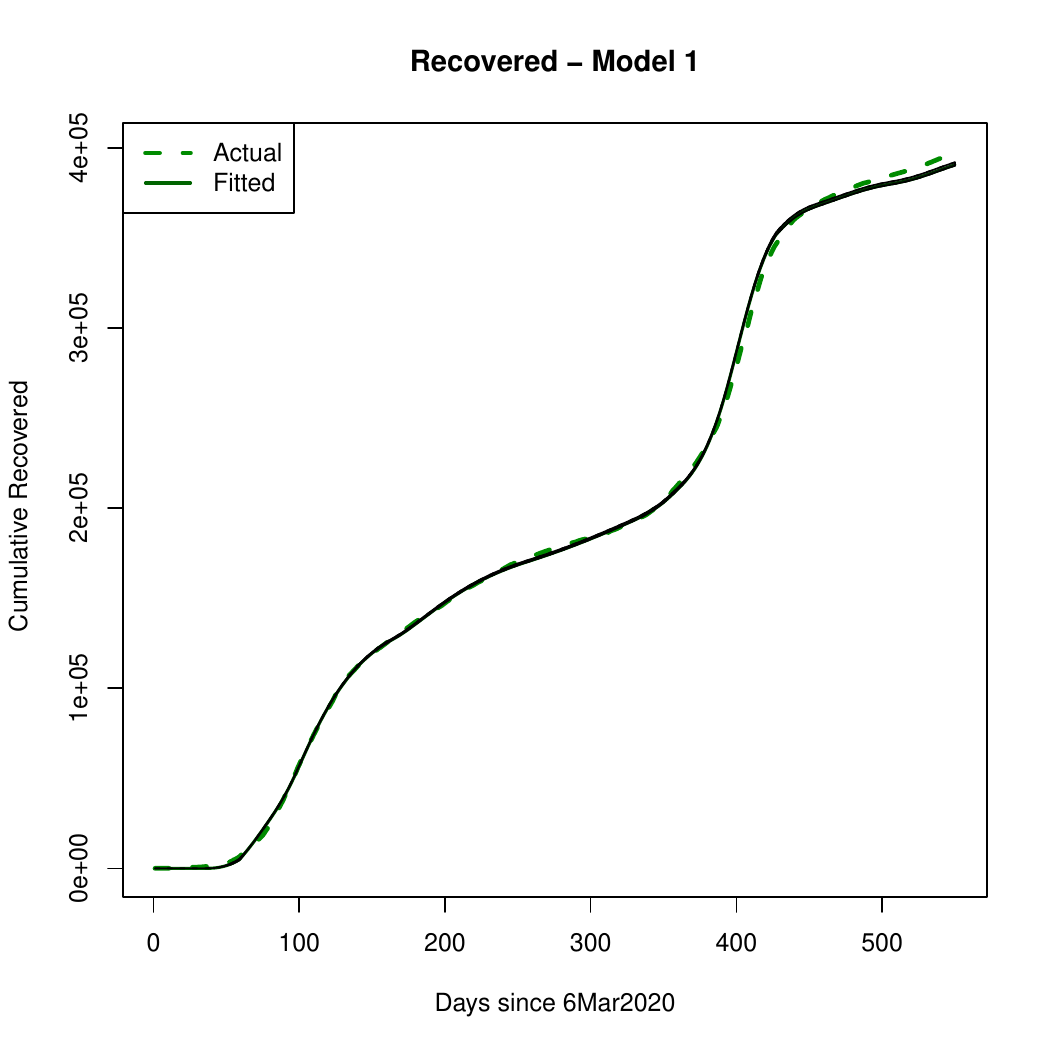}
     \caption{}
     \label{fig:b1}
 \end{subfigure}
 
 \medskip
 \begin{subfigure}{0.49\textwidth}
     \includegraphics[width=\textwidth]{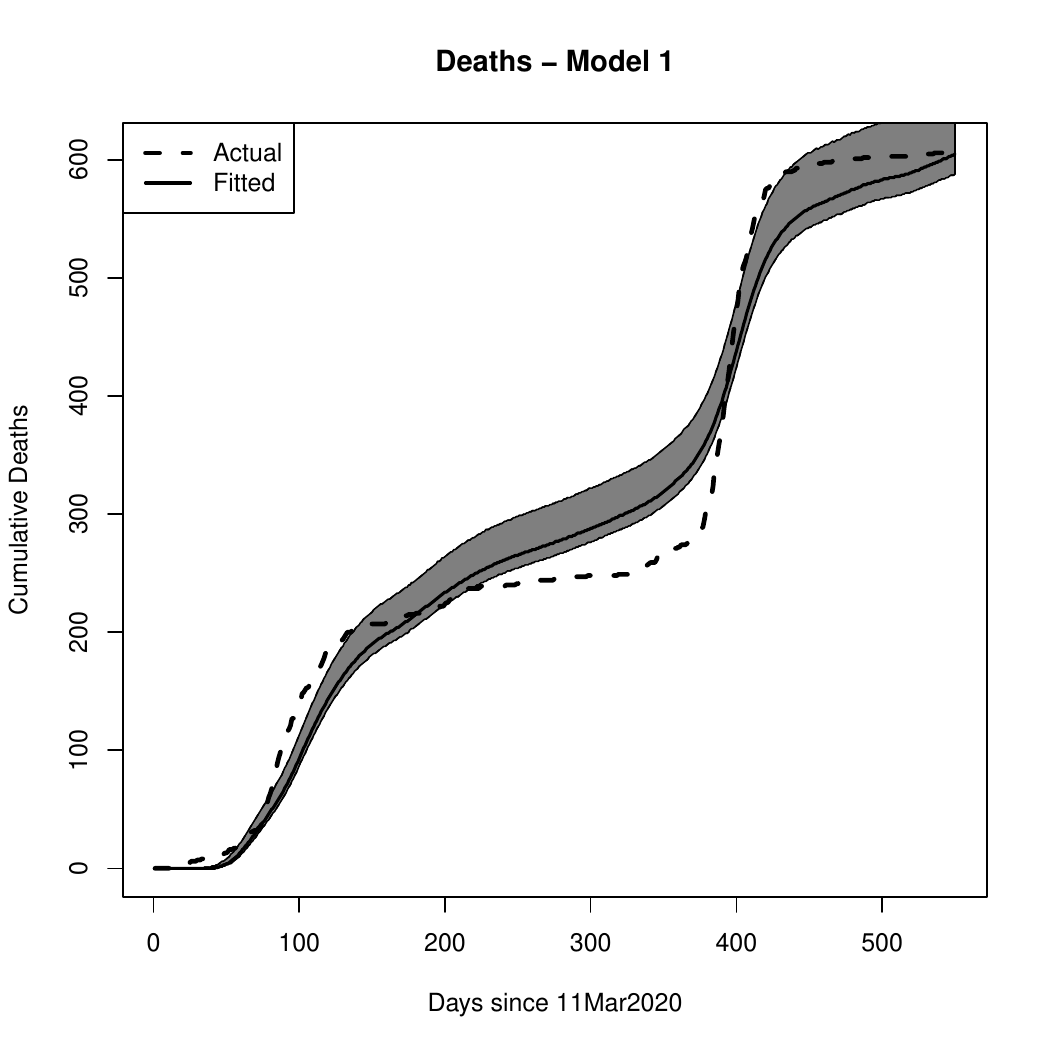}
     \caption{}
     \label{fig:c1}
 \end{subfigure}
 \hfill
 \begin{subfigure}{0.49\textwidth}
     \includegraphics[width=\textwidth]{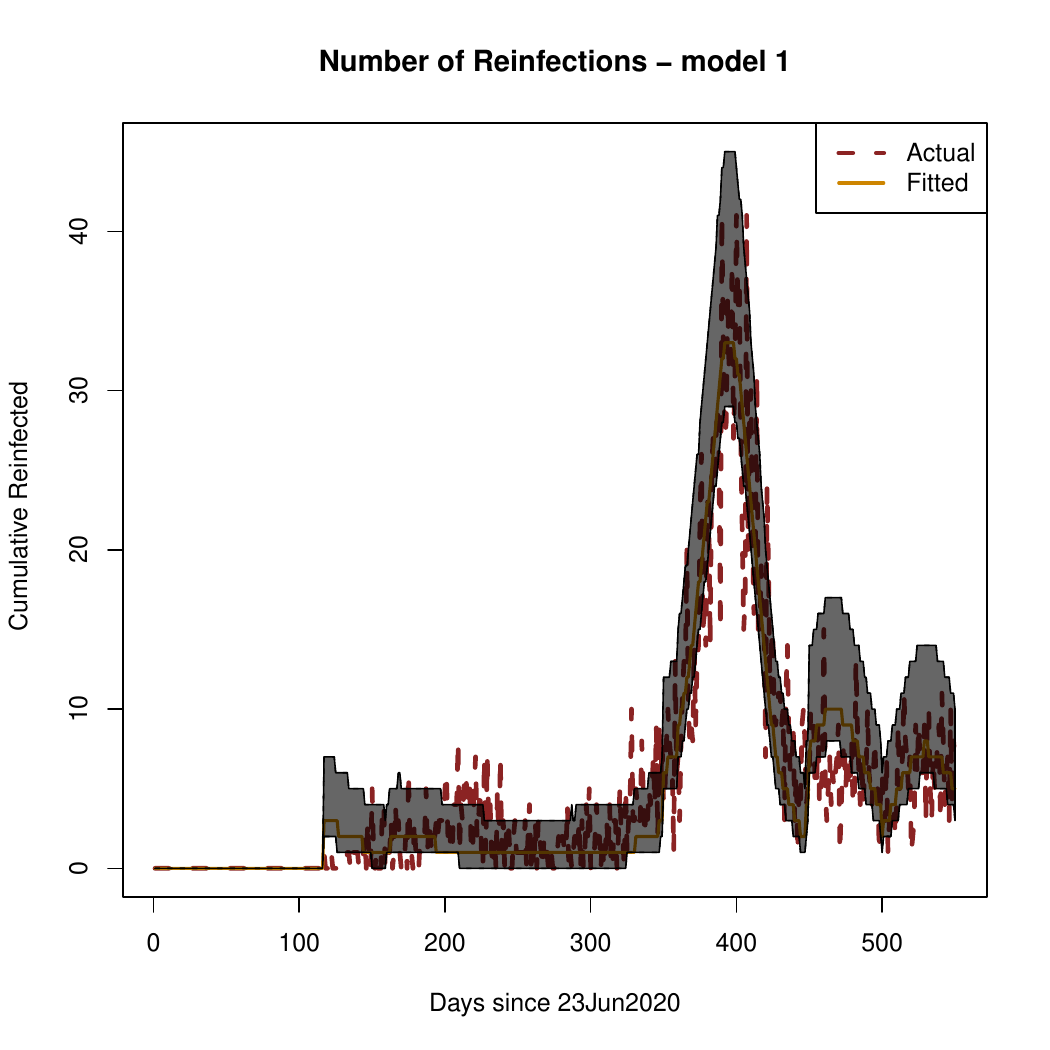}
     \caption{}
     \label{fig:d1}
 \end{subfigure}

 \medskip
 \begin{subfigure}{0.49\textwidth}
     \includegraphics[width=\textwidth]{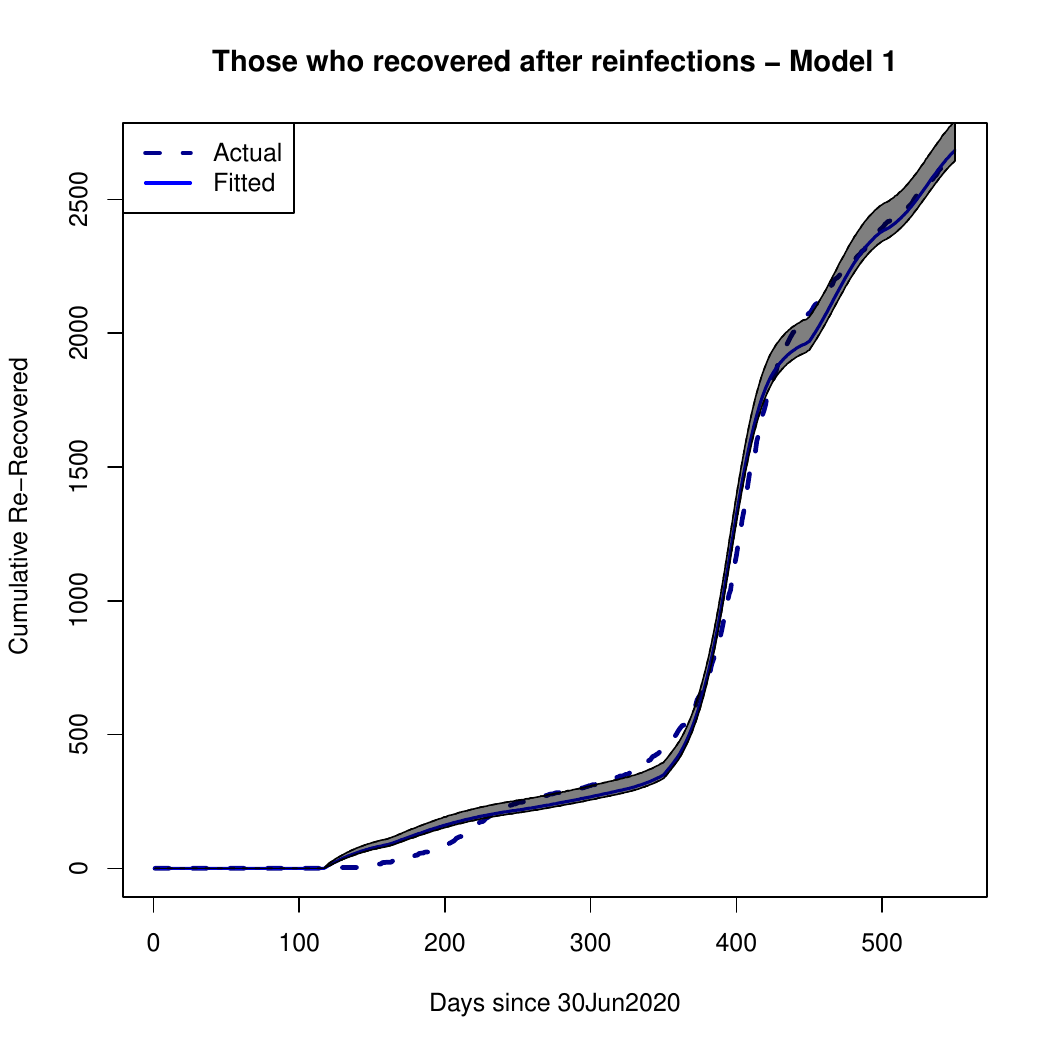}
     \caption{}
     \label{fig:e1}
 \end{subfigure}
 \hfill
 \begin{subfigure}{0.49\textwidth}
     \includegraphics[width=\textwidth]{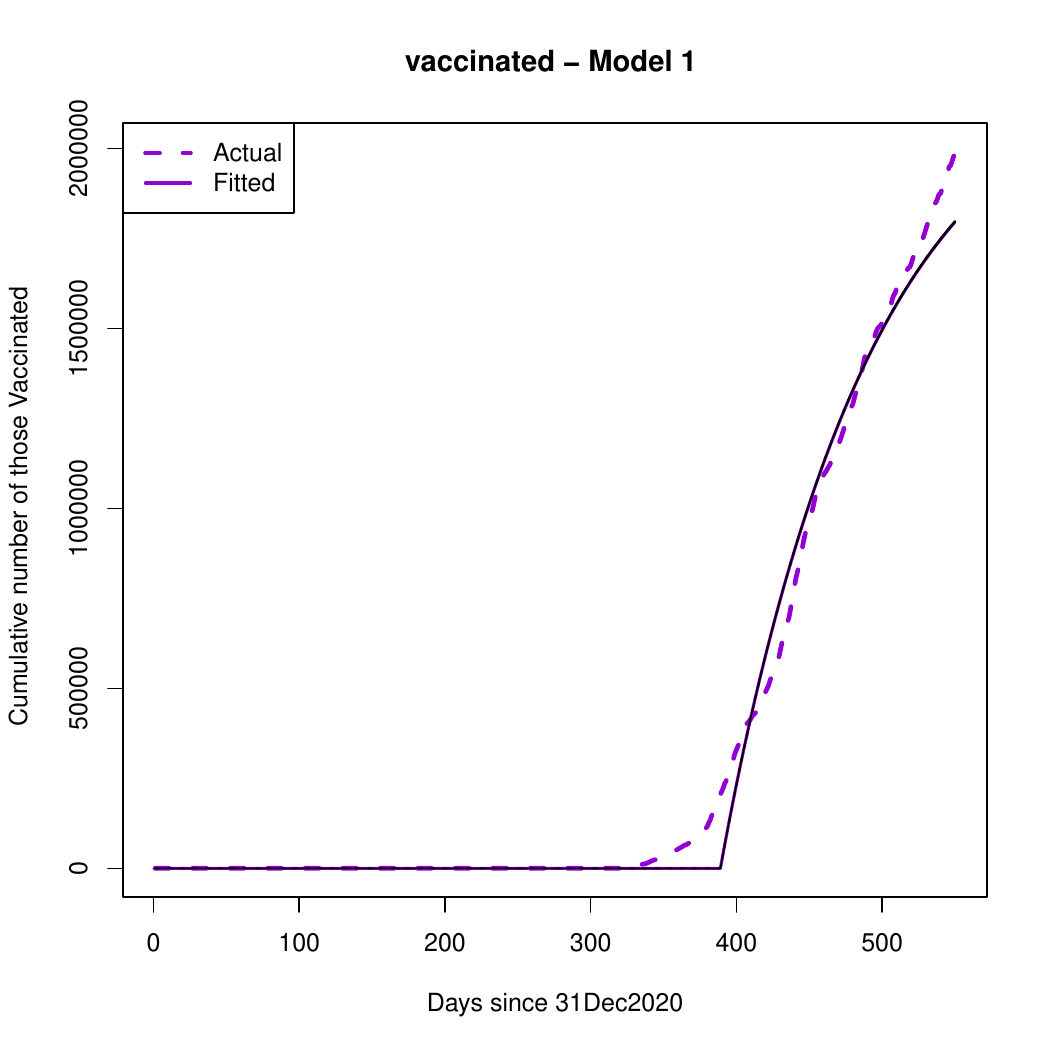}
     \caption{}
     \label{fig:f1}
 \end{subfigure}

 \caption{Plots of (a) Active Infections, (b) Cumulative Recovered, (c) Cumulative Deaths, (d) Number of Reinfections, (e) Cumulative Recovered after reinfections and (f) Cumulative number of Vaccinated for Model 1. The estimated quantiles ($0.025, 0.5$, and $0.975$) were calculated using $50,000$ samples from the posterior distribution.}
 \label{fit1}

\end{figure}

\begin{figure}[h]
 \begin{subfigure}{0.49\textwidth}
     \includegraphics[width=\textwidth]{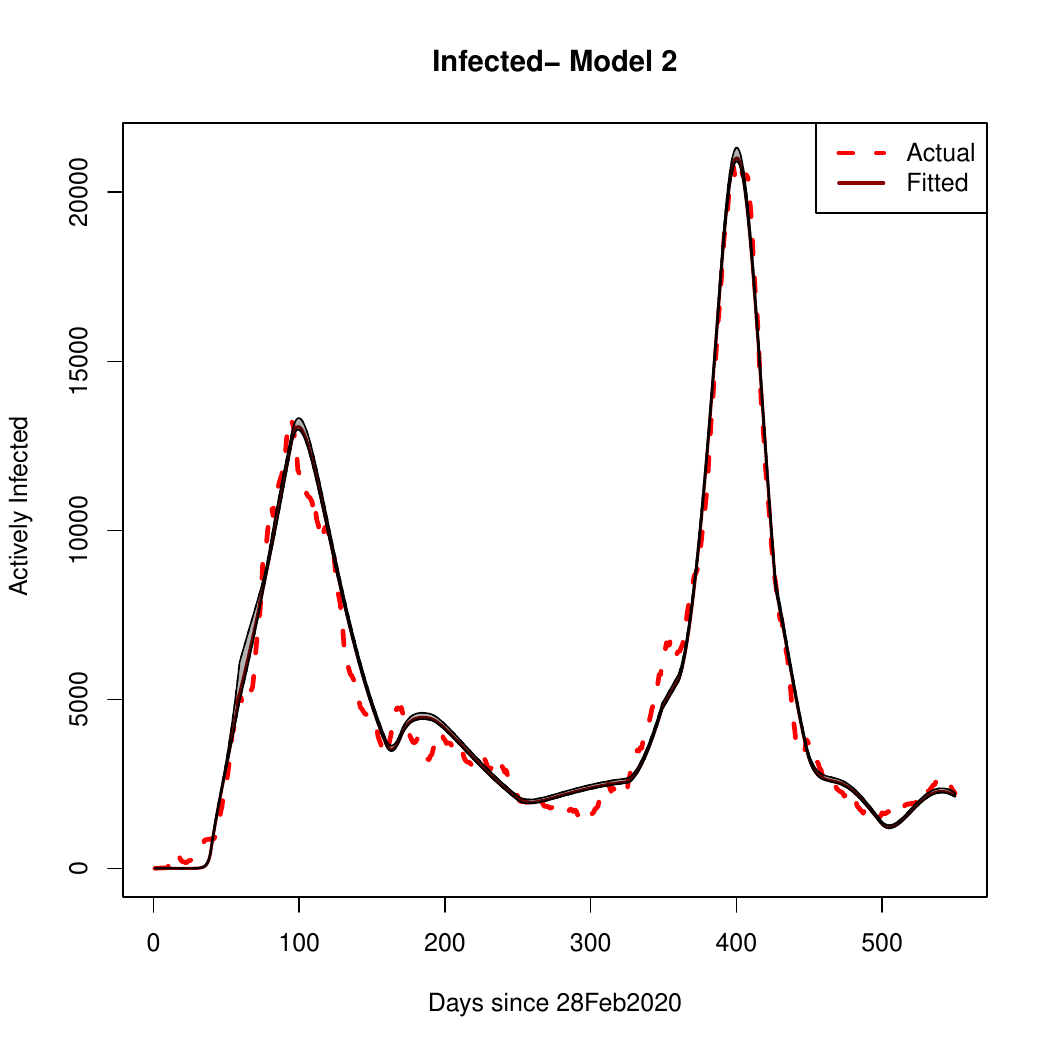}
     \caption{}
     \label{fig:a2}
 \end{subfigure}
 \hfill
 \begin{subfigure}{0.49\textwidth}
     \includegraphics[width=\textwidth]{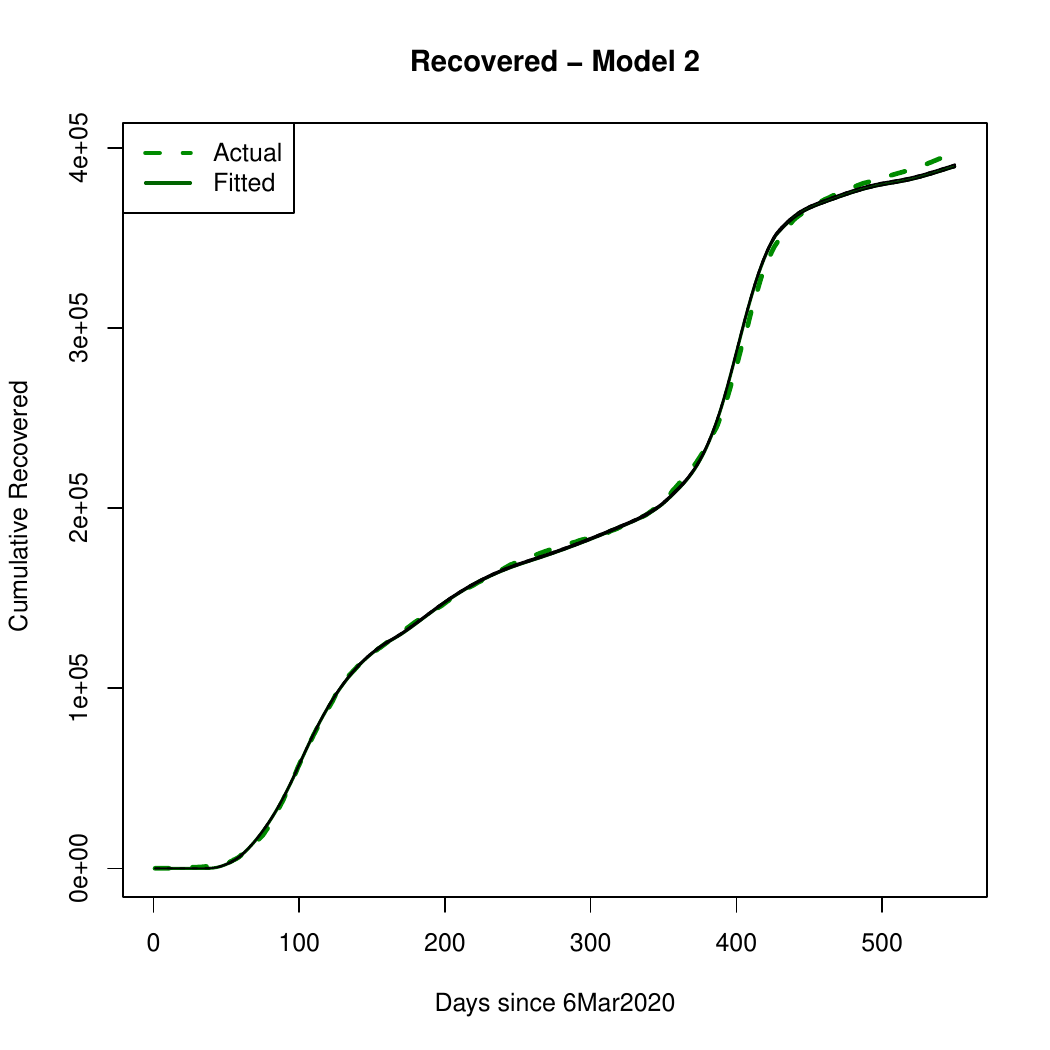}
     \caption{}
     \label{fig:b2}
 \end{subfigure}
 
 \medskip
 \begin{subfigure}{0.49\textwidth}
     \includegraphics[width=\textwidth]{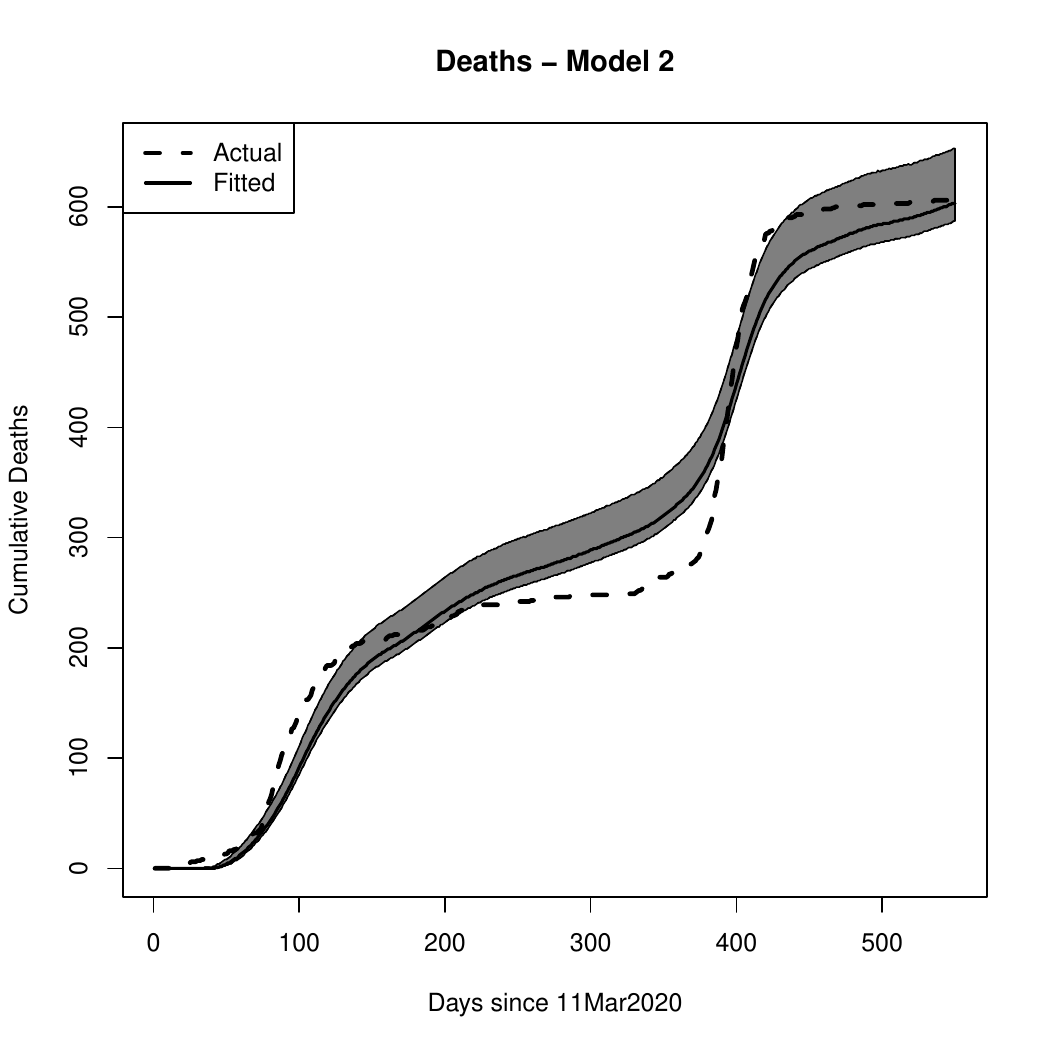}
     \caption{}
     \label{fig:c2}
 \end{subfigure}
 \hfill
 \begin{subfigure}{0.49\textwidth}
     \includegraphics[width=\textwidth]{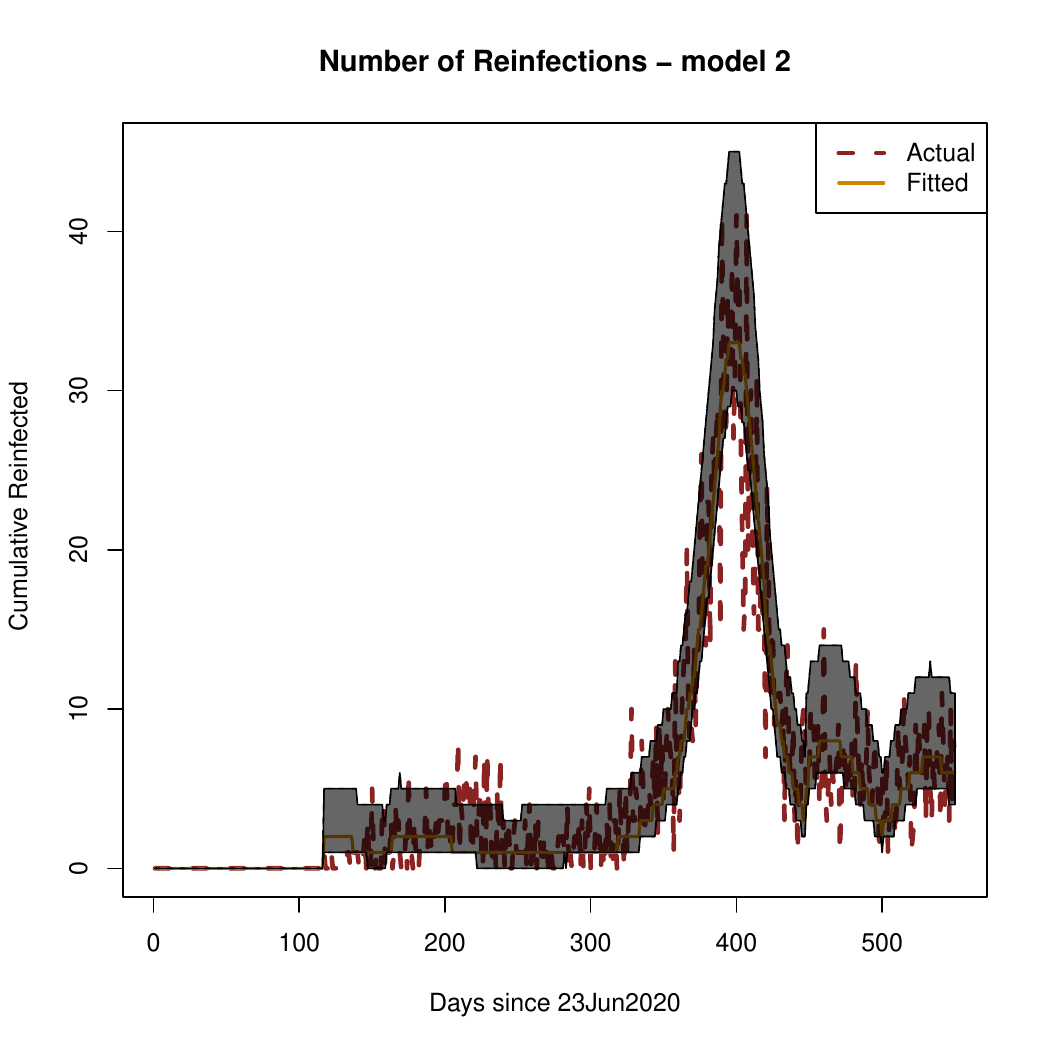}
     \caption{}
     \label{fig:d2}
 \end{subfigure}

 \medskip
 \begin{subfigure}{0.49\textwidth}
     \includegraphics[width=\textwidth]{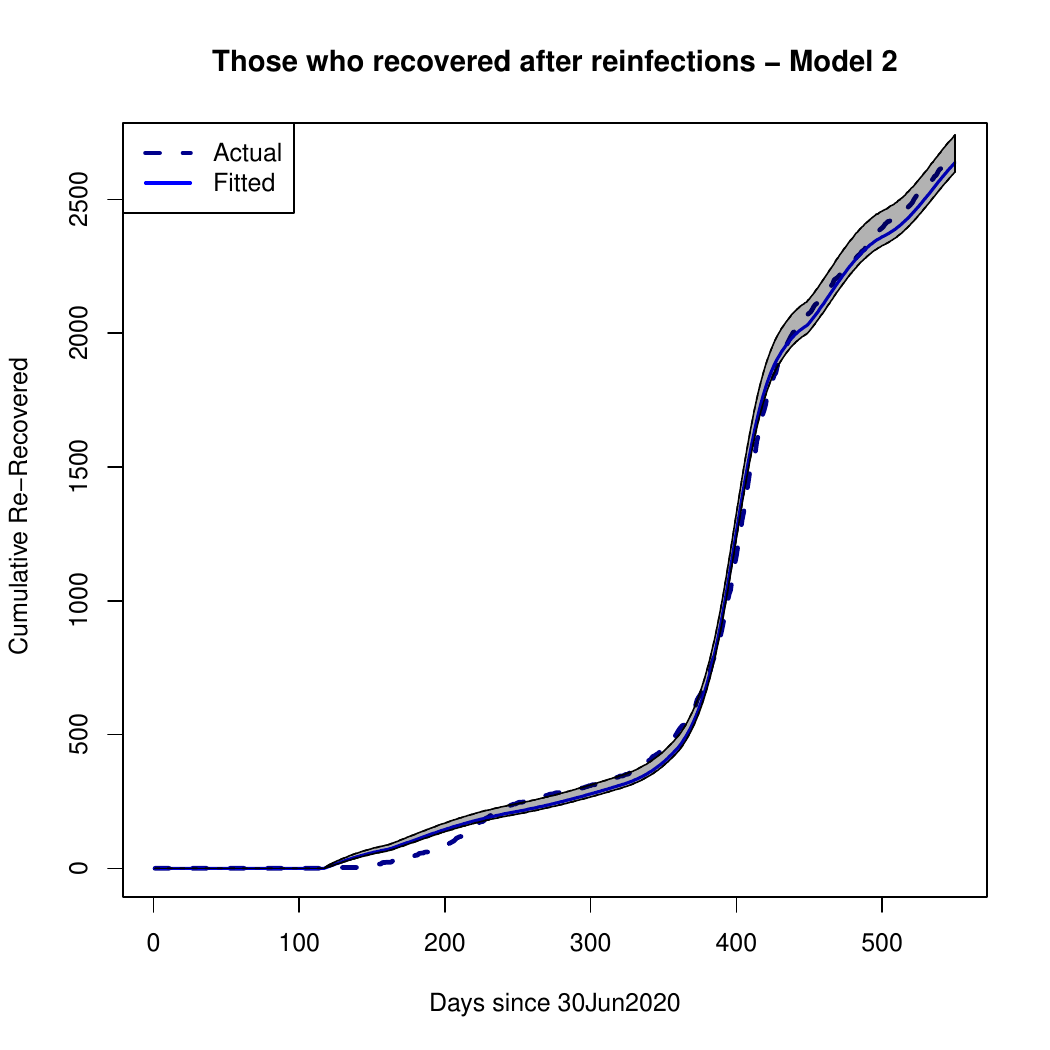}
     \caption{}
     \label{fig:e2}
 \end{subfigure}
 \hfill
 \begin{subfigure}{0.49\textwidth}
     \includegraphics[width=\textwidth]{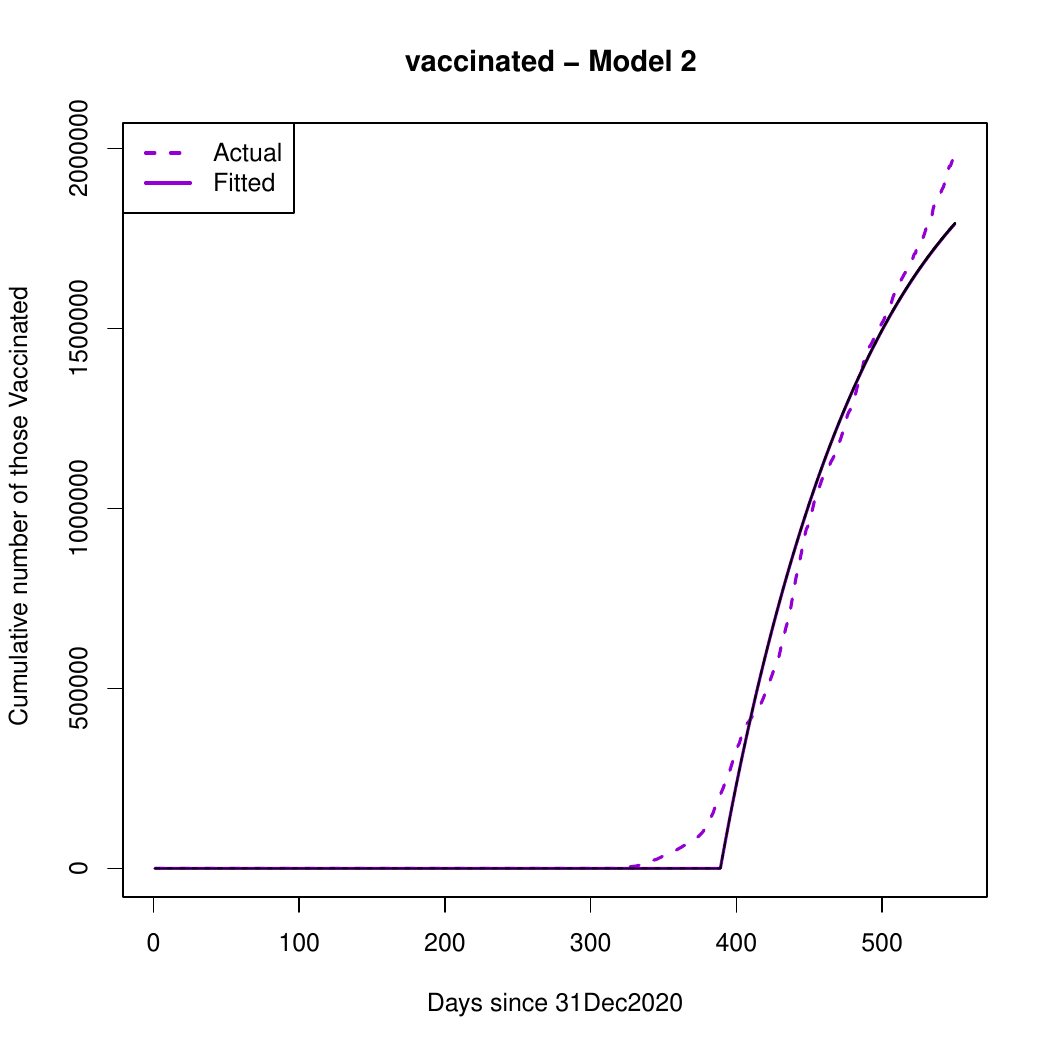}
     \caption{}
     \label{fig:f2}
 \end{subfigure}

 \caption{Plots of (a) Active Infections, (b) Cumulative Recovered, (c) Cumulative Deaths, (d) Number of Reinfections, (e) Cumulative Recovered after reinfections and (f) Cumulative number of Vaccinated for Model 2. The estimated quantiles ($0.025, 0.5$, and $0.975$) were calculated using $50,000$ samples from the posterior distribution.}
 \label{fit2}

\end{figure}

The initial conditions used for our analysis are presented in Table \ref{tbl:Init}. We also provide the results from fitting the two models, including estimated mean parameter, 2.5\% and 95\% quantiles, and pseudo-$R^2$ of the fitted models (shown in Table \ref{Tab1}) \cite{boone2023monitoring, amona2023incorporating}. The results were obtained by using the 50,000 samples generated from the posterior distribution. Model fits, along with estimated credible regions, are illustrated in Figures \ref{fit1} and \ref{fit2}, where actual data points are represented with dotted lines, and fitted values are denoted by solid lines. These models fit quite well, as is evident from the Psuedo-$R^2$ values in Table \ref{Tab1}. The rest of this section is organized as follows: Subsection \ref{scen12} examines vaccination timing, hospital overload, and various scenario analyses of the two models. Subsection \ref{modHel} delves into the similarity between the fitted density plots in the scenario analyses using Hellinger distance. The final subsection compares the plausibility of models \ref{eq:Sys1} and \ref{eq:Sys2} using the Bayes factor based on data from the State of Qatar. 

\begin{table}[ht!]
\begin{center}
\begin{tabular}{c c c  } \hline
Compartment & Initial conditions & Source \\
\hline
$S(0)= S_1(0)$ & $2,695,122$ & Population of Qatar as of December $30$, 2022 \\
$S_2(0)$ & $2695122/6$ & assumed \\
$V(0)$  & 0  & From the data \\
$E(0)$  & $5$ & From \cite{amona2023incorporating} \\
 $I(0)$  & 1 & From the data \\
 $R_E(0) = R_I(0)$ & 0 & From the data \\
 $I_I(0)$  & 0 & From the data \\
  $R_R(0)$ & 0 & From the data \\
$D(0)$ & 0 & From the data \\
\hline
\end{tabular}\caption{Initial conditions for models \ref{eq:Sys1} and \ref{eq:Sys2} }\label{tbl:Init}
\end{center}
\end{table}

\begin{table}[ht!]
\scalebox{0.59}{
\begin{tabular}{|c|c|l|l|l|l|} \hline
\textbf{Model}	&  \textbf{Parameter }	& \textbf{Description}	& \textbf{Mean Estimated Values}  &  \textbf{Quantiles ($0.025$ \& $0.975$)} &  \textbf{Psuedo}-\bm{$R^2$}\\
\hline \hline

& $\alpha_{0}$	& Transmission rate before intervention & $0.00286$ & ($0.00246, 0.00306$) & \\

& $\alpha_{1}$	& Transmission rate after the first intervention & $0.00084$ & ($0.00081, 0.00087$) & \\

& $\alpha_{2}$	& Transmission rate after the second intervention		& $0.00091$ & ($0.00089, 0.00095$) & \\

& $\alpha_{3}$	& Transmission rate after the third intervention		& $0.00076$ & ($0.00075, 0.00077$) & \\

& $\alpha_{4}$	& Transmission rate after the fourth intervention			& $0.00110$ & ($0.00109, 0.00111$) & \\

& $\alpha_{5}$	& Transmission rate after the fifth intervention		& $0.00089$ & ($0.00088, 0.00090$)& \\

& $\alpha_{6}$	& Transmission rate after the sixth intervention		& $0.00086$ & ($0.00085, 0.00088$) & \\

& $\alpha_{7}$	& Transmission rate after the seventh intervention		& $0.00097$ & ($0.00097, 0.00099$) & \\

& $\alpha_{8}$	& Transmission rate after the eighth intervention		&  $0.00112$ & ($0.00112, 0.00113$) & \\

& $\alpha_{9}$	& Transmission rate after the ninth intervention & $0.00132$ & ($0.00246, 0.00306$) & \\

& $\beta$		& Rate at which the exposed becomes infectious (Infectious rate	)	& $0.06988$ & ($0.06939 , 0.07075$) & \\

\textbf{Model 1} & $\gamma_{10}$	& Recovery rate from infections influenced by intervention		& $ 0.09097$ & ($0.05473, 0.10933$) &  $\bm{0.99176}$ \\

 & $\gamma_{11}$	& Recovery rate from infections influenced by intervention	& $0.13564$ & ($0.13499 , 0.13883$) & \\

& $\gamma_{12}$	& Recovery rate from infections influenced by intervention		& $0.14758$ & ($ 0.14192, 0.14847$) & \\

& $\gamma_{13}$	& Recovery rate from infections influenced by intervention	& $0.11333$ & ($0.11300 , 0.11370$) &  \\

& $\phi_{1}$        & Reinfection rate	influenced by intervention	& $0.00538$ & ($0.00528, 0.00543$)& \\

& $\phi_{2}$        & Reinfection rate	influenced by intervention	& $0.01187$ & ($0.01170, 0.01244$) &\\

& $\phi_{3}$        & Reinfection rate	influenced by intervention	& $0.02213$ & ($0.02118, 0.02312$) &\\

& $\phi_{4}$        & Reinfection rate	influenced by intervention	& $0.00740$ & ($0.00679, 0.00805$) & \\

& $\gamma_2$	& Recovery rate from Reinfections	& $0.14295$ & ($0.14295, 0.14295$)  & \\

& $\mu$        & Vaccination rates	& $0.00460$ & ($0.00000 , 0.00920$) & \\

& $\kappa$     & Vaccine efficacy	   & $0.94$ & ($0.94 , 0.94$)&  \\

& $\eta$        & Deaths rate	& $0.00021$ & ($0.00021 , 0.00021$) & \\

& $\zeta_1$      & Natural immunity waning rate	 & $ 0.00202$ & ($0.00001, 0.00046 $) & \\

& $\zeta_2$      & Immunity due to vaccination waning rate		& $0.00119$ & ($0.00017 0.00929 $)&  \\
\hline \hline
& $\alpha_{0}$	& Transmission rate before intervention		& $0.00314$ & ($0.00289, 0.00329$)&  \\

& $\alpha_{1}$	& Transmission rate after the first intervention		& $0.00089$ & ($0.00084 , 0.00093$) & \\

& $\alpha_{2}$	& Transmission rate after the second intervention		& $0.00089$ & ($0.00089, 0.00090$) & \\

& $\alpha_{3}$	& Transmission rate after the third intervention		& $0.00077$ & ($0.00076, 0.00079$) & \\

& $\alpha_{4}$	& Transmission rate after the fourth intervention		& $0.00110, $ & ($0.00109, 0.00110$) &  \\

& $\alpha_{5}$	& Transmission rate after the fifth intervention		& $0.00090$ & ($0.00088, 0.00091$) & \\

& $\alpha_{6}$	& Transmission rate after the sixth intervention		& $0.00087$ & ($0.00086, 0.00088$) &  \\

& $\alpha_{7}$	& Transmission rate after the seventh intervention		& $0.00098$ & ($0.00097, 0.00099$) & \\

& $\alpha_{8}$	& Transmission rate after the eighth intervention		& $00114$ & ($0.00113, 0.00115$) & \\

& $\alpha_{9}$	& Transmission rate after the ninth intervention	& $0.00135$ & ($0.00289, 0.00329$) & \\

& $\beta$		& Infectious rate		& $0.07270$ & ($0.06963 , 0.07616$) & \\
\textbf{Model 2}  & $\gamma_{10}$	& Recovery rate from infections influenced by intervention		& $0.11527$ & ($0.09406 , 0.12748$) &  $0.99160$\\

 & $\gamma_{11}$	& Recovery rate from infections influenced by intervention	& $0.13595$ & ($0.13537 , 0.13651$) & \\

& $\gamma_{12}$	& Recovery rate from infections influenced by intervention		& $0.15138$ & ($0.14984 , 0.15178$) & \\

& $\gamma_{13}$	& Recovery rate from infections influenced by intervention	& $0.11501$ & ($0.11460, 0.11544$) & \\

& $\phi_{1}$        & Reinfection rate	influenced by intervention	& $0.01885$ & ($0.01829 , 0.01945$) & \\

& $\phi_{2}$        & Reinfection rate	influenced by intervention			& $0.01978$ & ($0.01902 , 0.02071$) & \\

& $\phi_{3}$        & Reinfection rate	influenced by intervention			& $0.02315 $ & ($0.02159 , 0.02480$) & \\

& $\phi_{4}$        & Reinfection rate	influenced by intervention			& $0.01149$ & ($0.01041 , 0.01260$) & \\

& $\gamma_2$	& Recovery rate from Reinfections	& $0.14286$ & ($0.14286 , 0.14286$) & \\

& $\mu$        & Vaccination rates		& $0.00462$ & ($0.0000, 0.00925$) & \\

& $\kappa$     & Vaccine efficacy	    & $0.94$ & ($0.94 , 0.94$) & \\

& $\eta$        & Deaths rate	& $0.00021$ & ($0.00021, 0.00021$) & \\

& $\zeta_1$      & Natural immunity waning rate		& $0.05374$ &  ($0.00090 , 0.10646$) & \\

& $\zeta_2$      & Immunity due to vaccination waning rate	& $0.00359$ & ($ 0.00000, 0.00229$) & \\
\hline
\end{tabular}} \caption{Estimated $\bm{S_1}$EIRDV$\bm{S_2 I_I R_R}$ (Model 1) and SEIRDV$\bm{I_I R_R}$ (Model 2) model parameters. These parameter values are estimated using the 50,000 posterior samples. Times are in days, rates are individuals per day, except $\alpha, \zeta_1$ and $\zeta_2$ with rates in 10,000 individuals per day .\label{Tab1}}
\end{table}

\subsection{Scenario Analyses of Vaccine Efficacy and Timing in Hospital Overload using Models 1 and 2} \label{scen12}

At the onset of the COVID-19 pandemic, the Qatari government implemented various interventions to mitigate its spread. Among these measures, the stringent interventions proved notably effective in curtailing the rise of new COVID-19 cases, while the more lenient interventions appeared less effective \citep{amona2023incorporating}. Here we perform scenario analyses to assess the potential impact of not deploying the remaining stringent interventions during the COVID-19 pandemic in terms of number of days when new COVID-19 cases would surpass the total hospital bed capacity of 3134 beds for COVID-19 patients in Qatar, thereby completely overwhelming the medical system. We also aim to analyze any variations in this outcome across six potential scenarios. The vaccine was made available on day 380 (December 31, 2020), according to the available data, with an efficacy of $95 \%$ \citep{abu2021effectiveness}. However, the vaccine efficacy decreased as different variants emerged, leading us to choose the vaccine efficacy to be $94\%$ in our analysis. Early vaccination is chosen to be on day $200$. This is because we wanted to see the effect of the vaccine after the first wave recorded in the data (See red dotted lines in Figures \ref{fit1}(a) and \ref{fit2}(a)). Also, we chose the late vaccination to be on day $450$, after the second wave in the data. Thus, the scenarios studied in this analysis are namely; $94 \%$ vaccine efficacy (scenario 1), $100 \%$ vaccine efficacy (scenario 2), early vaccine (day $200$) with $94\%$ vaccine efficacy (scenario 3), late vaccine (day $450$) with $94\%$ vaccine efficacy (scenario 4), early vaccine with $100\%$ vaccine efficacy (scenario 5) and late vaccine with $100\%$ vaccine efficacy (scenario 6). In each scenario, we calculate the posterior predictive distribution of the daily count of infected individuals, subsequently determining the range of number of days for which the number of infected surpasses the predefined threshold of total available beds, based on 50,000 posterior samples.


Findings from Model 1 revealed that under partial interventions, when vaccine efficacy was 94\% (scenario 1), the number of days during which hospitals would have been overwhelmed ranged from 54 - 481 days. This signifies a period wherein the healthcare system would have exceeded its capacity, receiving more new patients than the established threshold, lasting between 54 and 481 days. A 100\% vaccine efficacy (scenario 2) would have narrowed the range down to 54 - 444 days. This underscores the significant reduction in overwhelmed days with a fully efficacious vaccine, even in the presence of partial interventions. Notably, introducing vaccines earlier (scenarios 3 and 5) did not substantially alter the overwhelmed period, which still remained at 54 - 444 days. However, if the vaccine introduction was delayed, the range considerably widened to 57 - 487 days. In summary, the analysis under Model 1 suggested that, under partial interventions, the Qatari government's provided hospital beds would never have sufficed without reaching critical capacity. Moreover, a 100\% efficacious vaccine or early administration, irrespective of efficacy, would have resulted in fewer overwhelmed days compared to delayed vaccine administration.

Similar to the findings in Model 1, Model 2's analysis also revealed that under the vaccine efficacy of 94\% (scenario 1), the hospitals' overwhelmed period ranged from 55 - 453 days, while a 100\% efficacy (scenario 2) narrowed it to 53 - 443 days, emphasizing the significance of highly efficacious vaccines in mitigating bed capacity challenges during partial interventions. The impact of early vaccine introduction under partial interventions was consistent across efficacy rates, maintaining a relatively stable overwhelmed period of 53 - 445 days under scenario 3 and 55 - 444 days under scenario 5. However, late vaccine introduction substantially extended the potential duration of hospital overload, with scenario 4 (94\% efficacy) ranging from 55 - 505 days and scenario 6 (100\% efficacy) from 53 - 513 days, surprisingly showing scenario 6 to be slightly less effective than scenario 4. This discrepancy raises questions about the model's realism and underscores the need for further exploration to identify an accurate model for capturing the real world dynamics of the pandemic.

\subsection{Hellinger Distance Analysis}\label{modHel}
\begin{figure}[h]
 \begin{subfigure}{0.49\textwidth}
     \includegraphics[width=\textwidth]{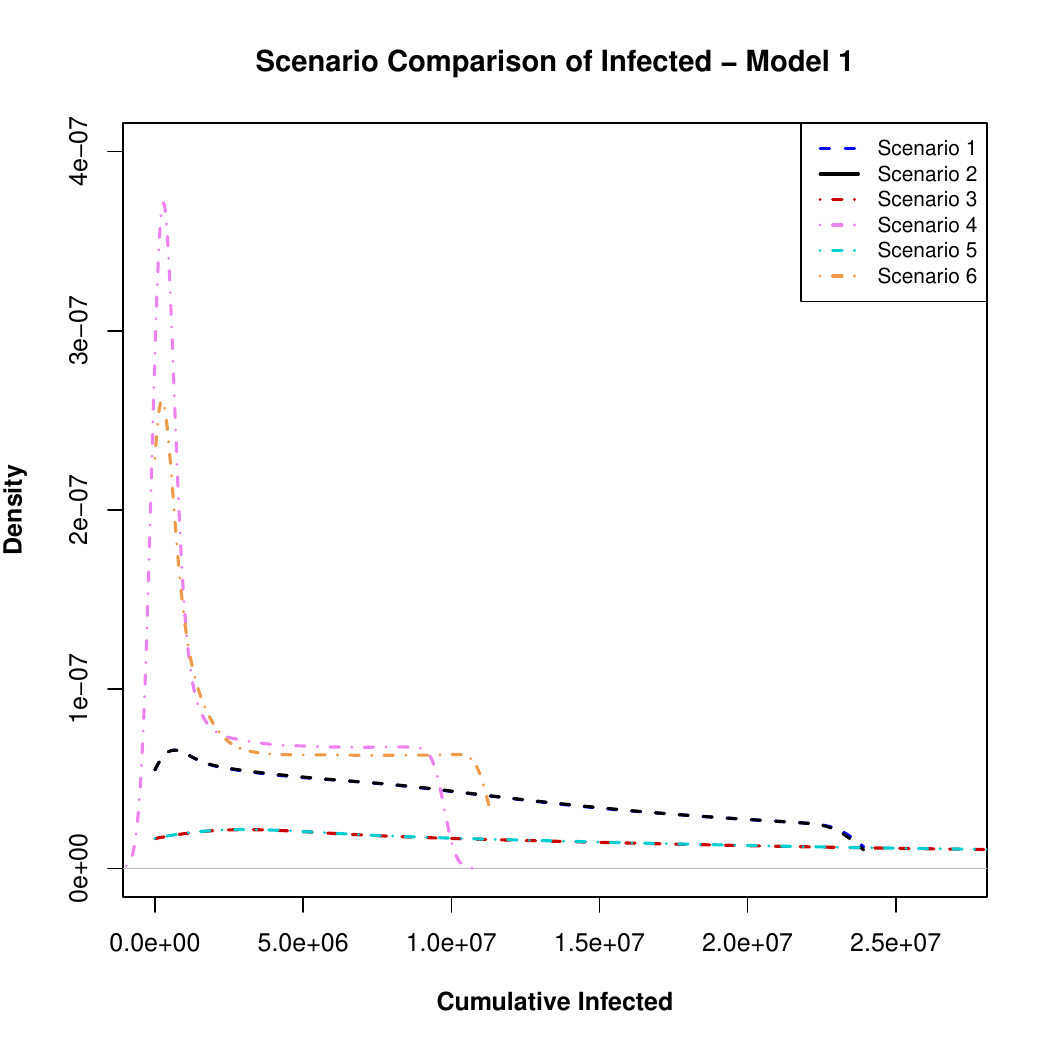}
     \caption{Comparison of PPDs for Cumulative Infected}
     \label{fig:a3}
 \end{subfigure}
 \hfill
 \begin{subfigure}{0.49\textwidth}
     \includegraphics[width=\textwidth]{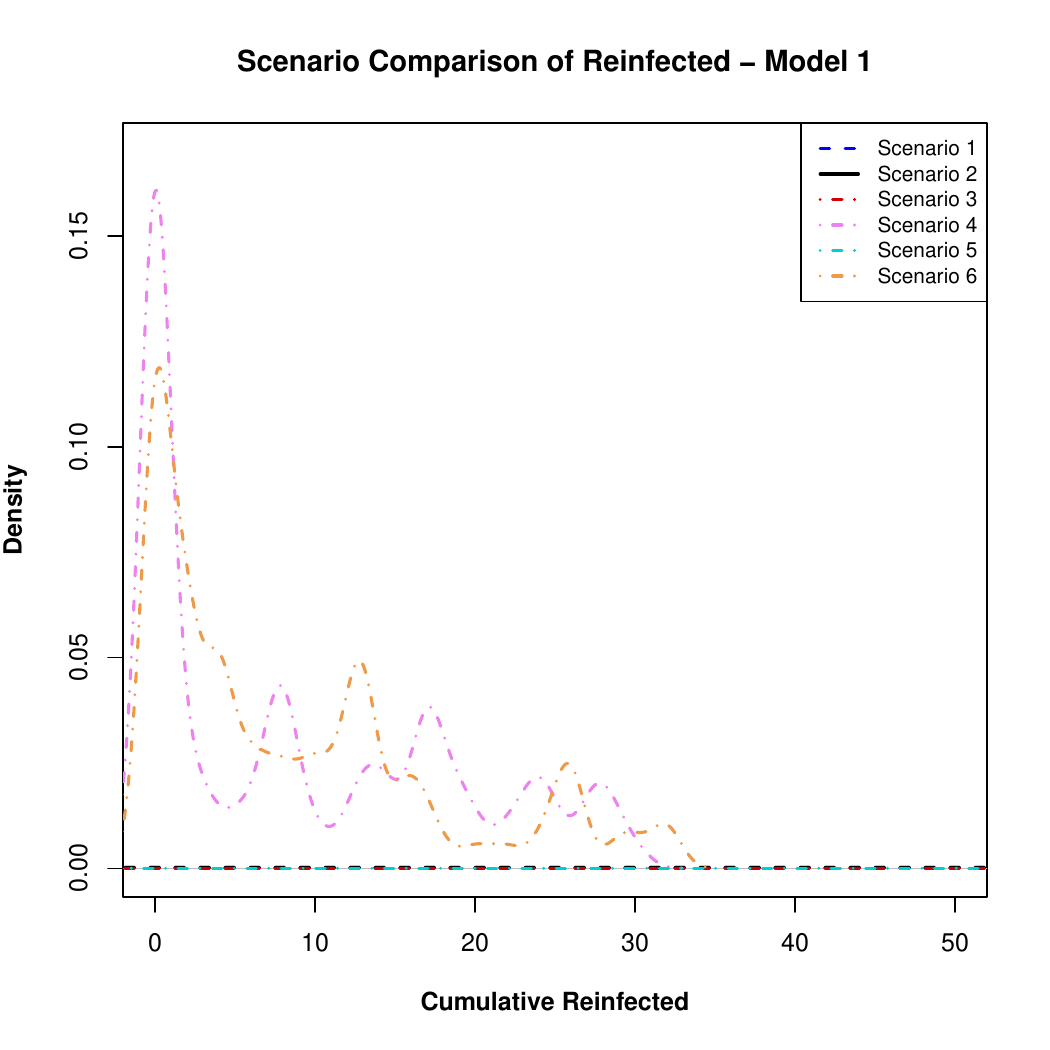}
     \caption{Comparison of PPDs for Cumulative Reinfected}
     \label{fig:b3}
 \end{subfigure}

  \medskip
  \centering
 \begin{subfigure}{0.49\textwidth}
     \includegraphics[width=\textwidth]{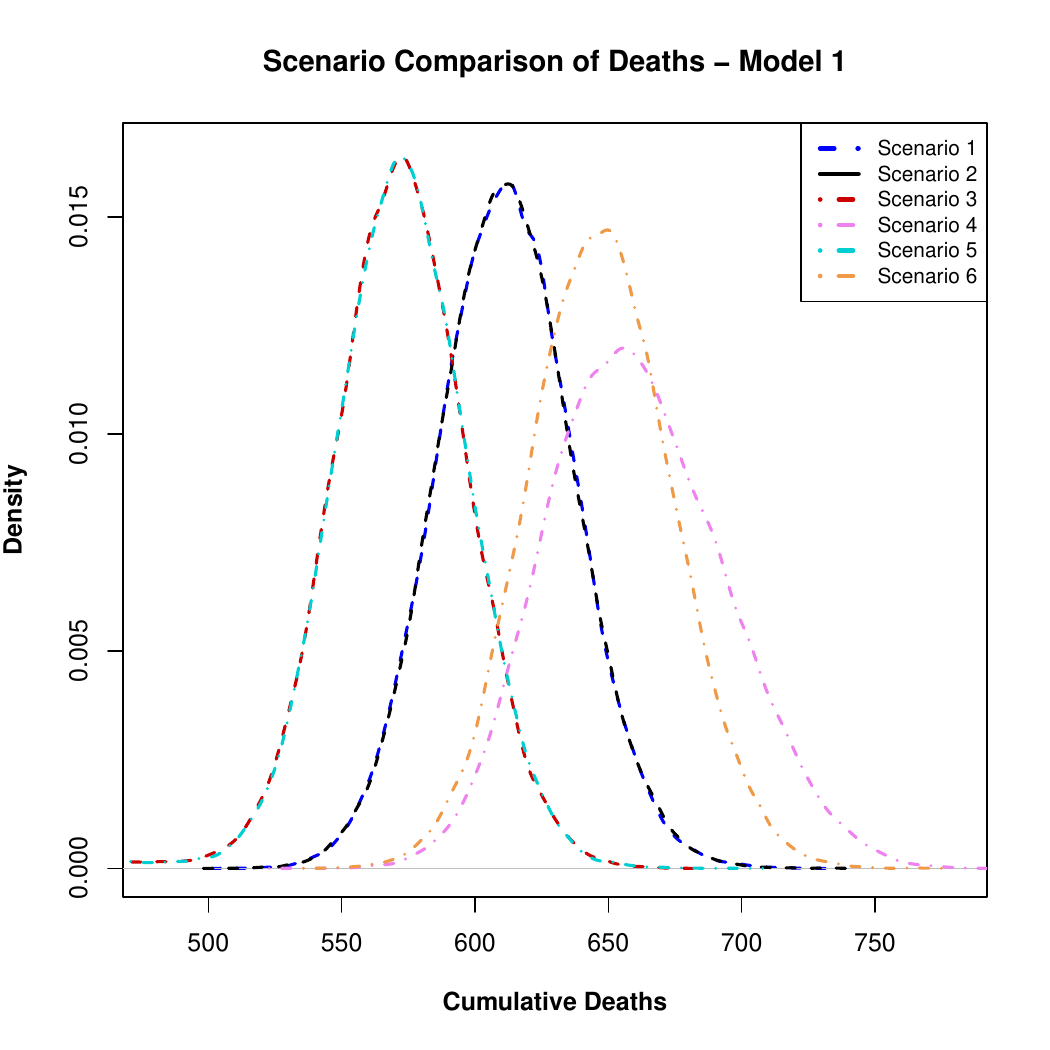}
     \caption{Comparison of PPDs for Cumulative Deaths}
     \label{fig:c3}
 \end{subfigure}
\caption{Plots of posterior predictive distributions (PPD) of (a) cumulative Infected, (b) Cumulative Reinfected and (c) Cumulative Deaths at day 540, obtained from Model 1. These were obtained using 50,000 samples from the posterior predictive distribution.}
 \label{scenarios.1}
\end{figure}

\begin{figure}[h]
 \begin{subfigure}{0.49\textwidth}
     \includegraphics[width=\textwidth]{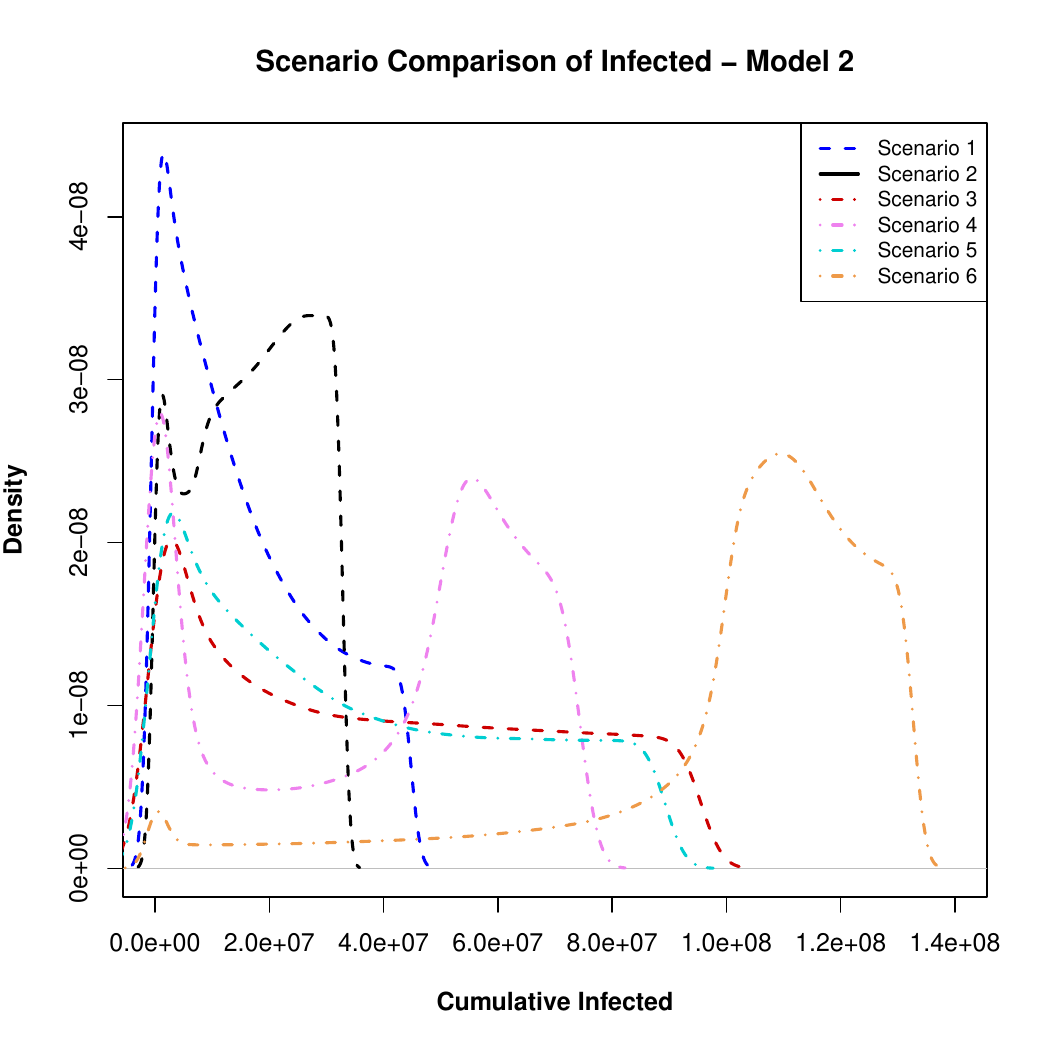}
     \caption{Comparison of PPDs for Cumulative Infected}
     \label{fig:a4}
 \end{subfigure}
 \hfill
 \begin{subfigure}{0.49\textwidth}
     \includegraphics[width=\textwidth]{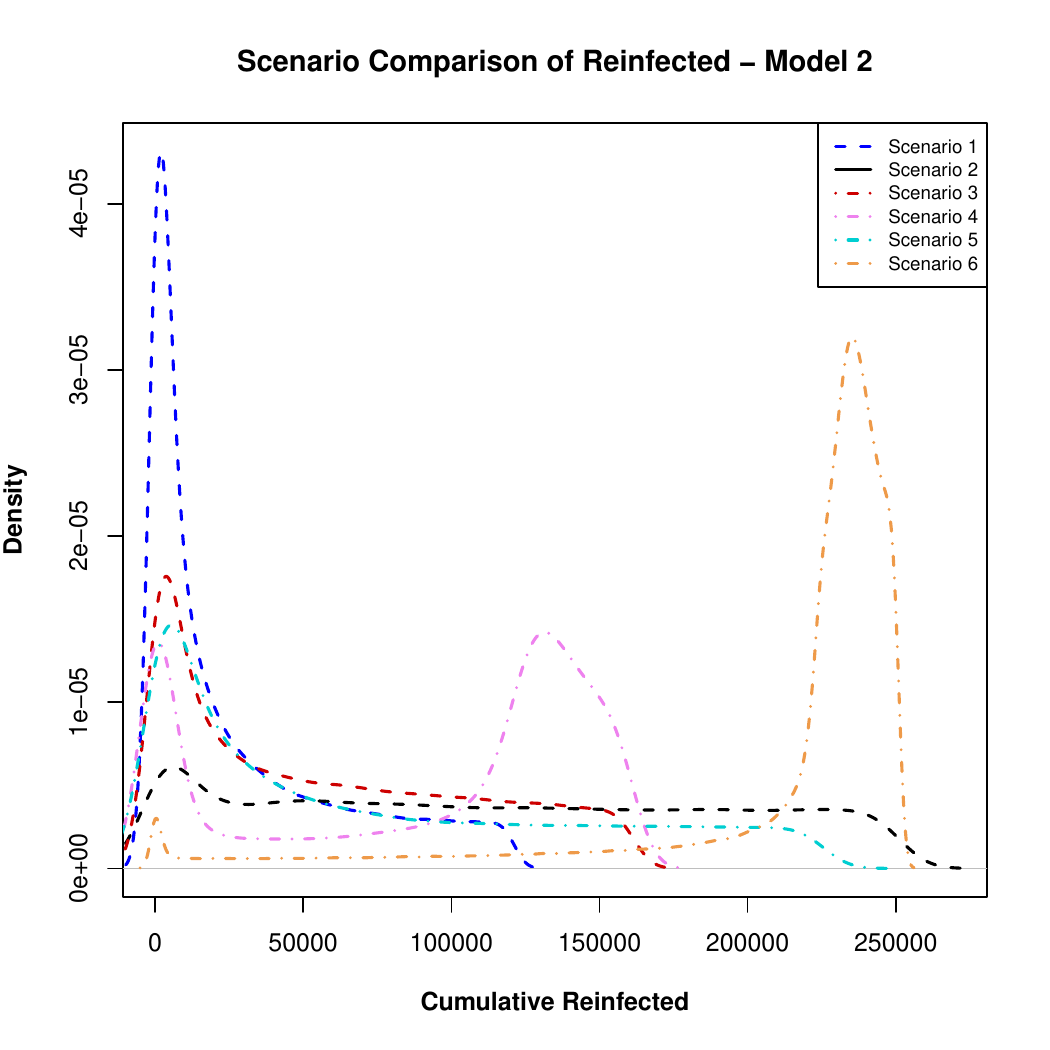}
     \caption{Comparison of PPDs for Cumulative Reinfected}
     \label{fig:b4}
 \end{subfigure}
 
 \medskip
 \centering
 \begin{subfigure}{0.49\textwidth}
     \includegraphics[width=\textwidth]{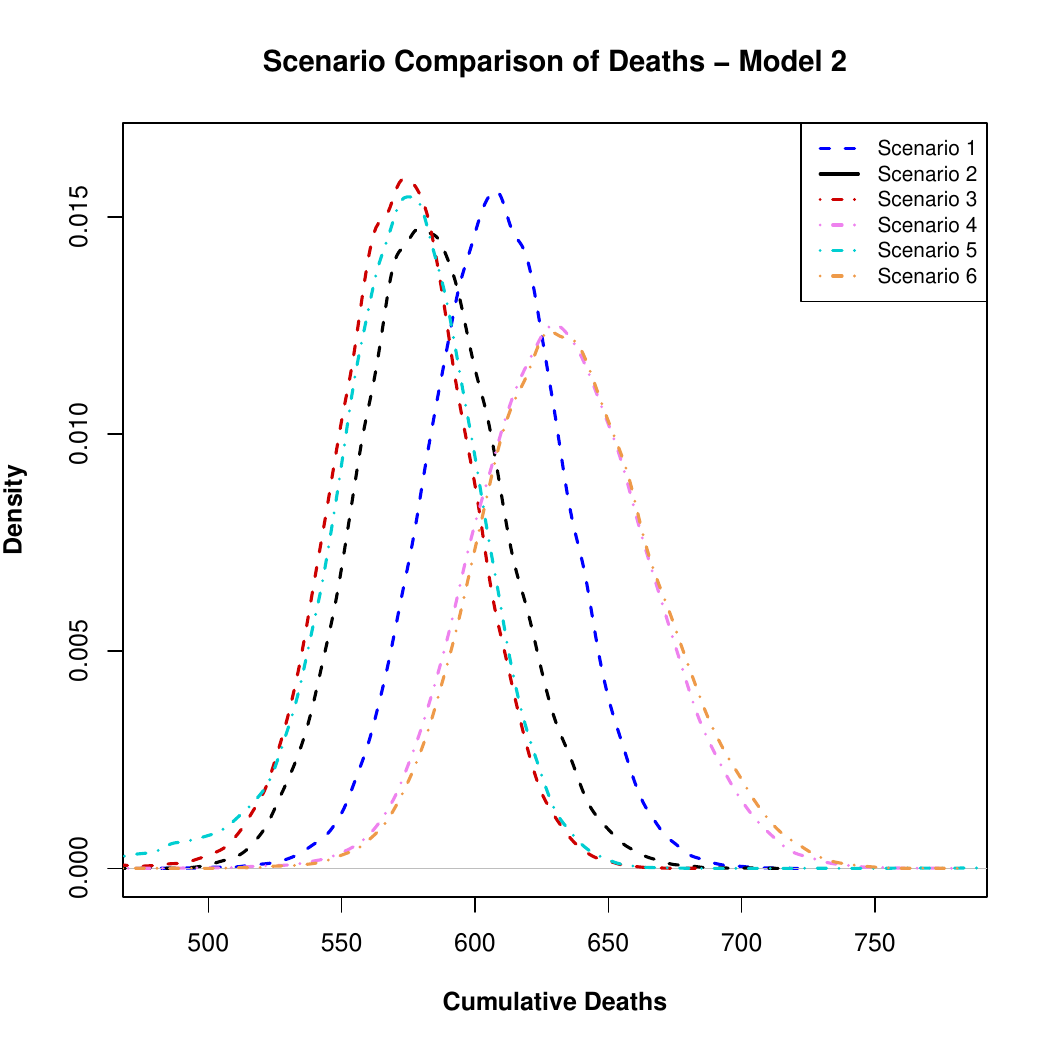}
     \caption{Comparison of PPDs for Cumulative Deaths}
     \label{fig:c4}
 \end{subfigure}
 \caption{Plots of posterior predictive distributions (PPDs) of (a) cumulative Infected, (b) cumulative Reinfected and (c) cumulative Deaths at day 540, obtained from Model 2. This was obtained using 50,000 samples from the posterior predictive distribution.}
 \label{scenarios.2}
\end{figure}

Under each scenario, we also computed the posterior predictive distributions for cumulative infections, reinfections, and deaths at day 540, which is 10 days before the conclusion of our study period on day 550. These distributions were smoothed using kernel density smoothing (KDE) with a Gaussian kernel and default bandwidth, and then graphed. This approach allowed us to clearly visualize the impact of varying vaccine efficacies and the timing of their availability on infections, reinfections, and deaths since the onset of the pandemic and the introduction of vaccines in Qatar. Figure \ref{scenarios.1} displays the smoothed posterior predictive distributions for cumulative infections, reinfections, and deaths at day 540 from Model \ref{eq:Sys1}, while Figure \ref{scenarios.2} illustrates the same for Model \ref{eq:Sys2}. Subsequently, we conducted Hellinger distance analysis on these smoothed densities to estimate the 'closeness' of the different scenarios.

In our examination of Model 1 posterior predictive distributions of cumulative infections at day 540 out of 550 in Figure \ref{scenarios.1}(a), scenarios involving the 94\% efficacious vaccine (scenario 1) and the 100\% efficacious vaccine (scenario 2) exhibited remarkably high similarity, indicated by a low Hellinger Distance of $0.000020$, indicating that the pursuit of an ideal vaccine may not have significantly influenced the effective control of cases within the system. Conversely, dissimilarity was substantial when comparing scenarios 1 and 3 (early vaccine with 94\% efficacy), with a Hellinger distance of $0.340625$. Late vaccinations (scenarios 4 and 6) resulted in higher cumulative infections and demonstrated considerable dissimilarity, marked by a Hellinger distance of $0.034114$, with $100\%$ efficacy being more effective than $94\%$ efficacy. 

Shifting to reinfections posterior predictive density plots from Model 1 in Figure \ref{scenarios.1}(b), scenarios 1 and 2 showed significant similarity (Hellinger Distance: $0.000179$), while dissimilarity was pronounced between scenarios 1 and 3 (Hellinger Distance: $0.375517$). Scenarios 3 and 5 displayed substantial similarity (Hellinger Distance: $0.000012$) but scenarios 4 and 6 showed some dissimilarity (Hellinger Distance: $0.033001$). Also, cumulative reinfections were likely to be around $35$ on day $540$ when the vaccine was administered late (scenarios 4 and 6). The most substantial dissimilarity emerged when comparing Scenarios 5 and 6, signified by a Hellinger Distance of $0.531083$.

The Model 1 posterior predictive cumulative death density plots in Figure \ref{scenarios.1}(c) revealed that early vaccination (scenarios 3 and 5) would have resulted in lower number of cumulative deaths compared to the observed outcomes (scenarios 1 and 2). Late vaccination (scenarios 4 and 6) would have produced a much higher cumulative death toll and vaccine efficacy mattered in this case. Significant similarity was seen between scenarios 1 and 2 (Hellinger Distance: $0.000091$) and scenarios 3 and 5 (Hellinger Distance: $0.000278$), while moderate to substantial dissimilarities were observed between scenarios 4 and 6 (Hellinger Distance: $0.031157$) and scenarios 5 and 6 (Hellinger Distance: $0.644322$) respectively. These findings show that lower infections lead to fewer deaths, while higher infections lead to higher number of deaths, thereby mimicking real-world dynamics.

In line with our analysis for Model 1, we computed Hellinger distances for the posterior predictive infected density plots in Model 2, as displayed in Figure \ref{scenarios.2}(a). Unlike Model 1, scenario 1 (94\% efficacious vaccine) and scenario 2 (100\% efficacious vaccine) demonstrated remarkable dissimilarity (Hellinger Distance: $0.500488$). Comparison between scenario 1 with scenario 3 (early vaccine with 94\% efficacy) revealed a moderate degree of dissimilarity (Hellinger Distance: $0.242202$), and that between scenario 3 and scenario 5 (early vaccine at 100\% efficacy) displayed moderate similarity as well (Hellinger Distance: $0.013369$). Conversely, the comparison between scenario 4 (late vaccine at 94\% efficacy) and scenario 6 (late vaccine at 100\% efficacy) revealed the most significant dissimilarity (Hellinger Distance: $0.616759$). The same trend was also observed between scenarios 5 and 6 (Hellinger Distance: $0.558894$). Overall, the efficacy of the vaccine appeared to be crucial in Model 2, with a clear preference for a $100\%$ efficacious vaccine over a $94\%$ efficacious one, irrespective of timing of roll out.

Transitioning to reinfections posterior probability density plots in Figure \ref{scenarios.2}(b), scenarios 1 and 2 exhibited substantial dissimilarity (Hellinger Distance: $0.298233$). Similarly, comparing scenarios 1 and 3, we observed substantial dissimilarity (Hellinger Distance: $0.102579$). Conversely, scenario 3 and scenario 5 displayed a lesser degree of dissimilarity (Hellinger Distance: $0.07729328$). Additionally, scenarios 4 and 6 were found to have significant dissimilarity (Hellinger Distance: $0.5072856$), while scenarios 5 and 6 were moderately dissimilar (Hellinger Distance: $0.3656723$). Thus early vaccination was still preferred, as the probability of excessively high reinfections became improbable in these scenarios.

For the cumulative posterior predictive distributions of death in Figure \ref{scenarios.2}(c), our analysis indicated moderate similarity between scenarios 1 and 2 (Hellinger distance: $0.094946$). Considerable similarity was also observed between scenario 3 and 5 (Hellinger distance: $0.006788$) and scenarios 4 and 6 (Hellinger distance: $0.001163$). The most significant dissimilarity was observed when comparing scenarios 5 and 6 (Hellinger Distance: $0.440002$). These results once again advocated for early vaccination, with the higher efficacious vaccine resulting in lower cumulative deaths. 

\subsection{Model Comparison using Bayes factor}
In our study, we have introduced two distinct models, each incorporating unique underlying dynamics, aiming to identify the model better suited to explain the observed data. Model 1 introduces a secondary compartment ($S_2$) to account for individuals with recovered natural immunity. Additionally, it considers the waning of vaccine-induced immunity, while individuals transition to the secondary susceptible compartment at varied rates. Conversely, Model 2 posits that a second susceptible compartment is unnecessary. Our focus is on comparing the performance of these models through the Bayes factor metric, detailed in Subsection \ref{bayes}, to determine their relative likelihood of explaining the observed data. Based on the data, the Bayes factor ($BF_{1,2}$) was calculated to be $50013.35$, which is greater than 1. Hence we conclude that Model 1 is more probable (a better fit for the data) than Model 2. This is also corroborated by the higher Psuedo-$R^2$ value of Model 1 compared to Model 2. Thus, the Bayes factor analysis reveals that Model 1, incorporating a secondary compartment ($S_2$) to accommodate individuals with recovered natural immunity and addressing the waning of vaccine-induced immunity with varied transition rates to the secondary susceptible compartment, is deemed more appropriate for representing the dynamics of the COVID-19 pandemic in the State of Qatar.


\section{Discussion}
In the realm of mathematical modeling, we are often tasked with studying the intricate dynamics of real-world phenomena. To this end, we implement models which offer distinct perspectives on complex systems. In this study, we introduce two such models, each providing a unique lens through which we explore an important question: How do vaccination, reinfection, and partial interventions impact the course of infectious diseases? We delve into these models with a clear goal: to understand which one aligns most closely with real-world observations. Additionally, we explore how changes in vaccine effectiveness and time of vaccine deployment impact the patterns of infections, reinfections, and deaths in these models. We also carefully examine the accuracy of certain assumptions in our models, especially the idea that people recovering from initial infections may have varying susceptibilities to reinfection.

To address these fundamental questions, we undertake a comparative analysis of our mathematical models by performing model selection using Bayes factor. Our analysis shows that based on real-life COVID-19 data from the State of Qatar, the model assuming varying susceptibilities among individuals recovering from their initial infections (Model 1) is significantly more probable than the simpler model, which assumes uniform susceptibility among all individuals (Model 2). Moreover, our approach of employing Bayes factor for model selection in the context of disease modeling is novel and, to the best of our knowledge, has not been explored before in literature.

Our research can also serve as a valuable tool for decision-makers grappling with the formulation of challenging public health policies during a pandemic. It is evident from Figure \ref{scenarios.1} (a-c)that, regardless of the vaccine's effectiveness, getting vaccinated late is not recommended. This late vaccination increases the chances of more people getting infected, reinfected, and more deaths. The timing of vaccine roll out also matters; our results show that introducing it earlier saved lives and reduced infections, although the overall decrease wasn't as significant as with early vaccination. Therefore, in the presence of partial interventions, early vaccination (scenarios 3 and 5) would be the most effective strategy in reducing active infections, reinfections, and deaths. If the vaccine had been given earlier, we could have saved more than the 50 lives, as mentioned in the study by \citep{amona2023incorporating}, and the number of infectious individuals would have decreased significantly compared to the reduction observed when the vaccine was introduced much later. As a result, our research advocates for the implementation of early vaccination protocols in the event of a future pandemic.

\vspace{6pt} 


\section*{Acknowledgements} 
\noindent The authors would like to thank Virginia Commonwealth University in Qatar and Qatar Foundation for supporting this work through the VCU Qatar Mathematical Data Sciences Lab.



\authorcontributions{Conceptualization, E.A., E.B. and R.G.; methodology, E.A., I.S. and E.B.; software, E.A.; validation, E.A. and I.S.; formal analysis, E.A.; investigation, E.A.; resources, E.B. and R.G.; data curation, E.A.; writing---original draft preparation, E.A. and I.S.; visualization, E.A.; supervision, I.S. and E.B.; project administration, I.S, E.B. and R.G.; funding acquisition, R.G. All authors have read and agreed to the published version of the manuscript.}



\dataavailability{The dataset and code used for this project can be found in \url{https://github.com/elizabethamona/VaccinationTiming}. The data includes  daily cumulative number of confirmed infections, cumulative number of recovered and the cumulative number of deaths for the State of Qatar starting on February 28, 2020. It also includes number daily reinfections and re-recovered for Qatar starting from June 23, 2020, and the number of vaccinated individuals in the country from December 31 2020.} 

\conflictsofinterest{The authors declare no conflict of interest.} \label{data}

\reftitle{References}
\externalbibliography{yes}
\bibliography{mybibliograhy.bib}

\end{document}